\documentclass[onecolumn,english,graybox]{svmult}
\pdfoutput=1
\usepackage[T1]{fontenc}
\usepackage[latin9]{inputenc}
\usepackage{amsmath}
\usepackage{graphicx}
\usepackage{amssymb}
\usepackage{esint}

\usepackage{cite}
\usepackage{url}

\renewcommand{\Re}{\operatorname{Re}}
\renewcommand{\Im}{\operatorname{Im}}
\newcommand{\tr}{\operatorname{tr}}

\usepackage{babel}

\begin{document}

\title{Numerical methods for computing Casimir interactions}

\author{Steven G. Johnson}

\institute{Steven G. Johnson \at Department of Mathematics, Massachusetts Institute
of Technology, Cambridge MA 02139 \email{stevenj@math.mit.edu}}

\maketitle

\begin{abstract}{}
We review several different approaches for computing
Casimir forces and related fluctuation-induced interactions between
bodies of arbitrary shapes and materials.  The relationships between
this problem and well known computational techniques from classical
electromagnetism are emphasized.  We also review the basic principles
of standard computational methods, categorizing them according to
three criteria---choice of problem, basis, and solution
technique---that can be used to classify proposals for the Casimir
problem as well.  In this way, mature classical methods can be
exploited to model Casimir physics, with a few important modifications.
\end{abstract}

\section{Introduction}

Thanks to the ubiquity of powerful, general-purpose computers, large-scale
numerical calculations have become an important part of every field
of science and engineering, enabling quantitative predictions, analysis,
and design of ever more complex systems. There are a wide variety
of different approaches to such calculations, and there is no single
{}``best'' method for all circumstances---not only are some methods
better suited to particular situations than to others, but there are
also often severe trade-offs between generality/simplicity and theoretical
efficiency. Even in relatively mature areas like computational classical
electromagnetism (EM), a variety of techniques spanning a broad range
of sophistication and generality remain in widespread use (and new
variations are continually developed)~\cite{Rao99,Taflove00,Volakis01,chew01,Jin02,Yasumoto05,Zhu06,JoannopoulosJo08-book}.
Semi-analytical approaches also remain important, especially perturbative
techniques to decompose problems containing widely differing length
scales (the most challenging situation for brute-force numerics).
Nevertheless, many commonalities and guiding principles can be identified
that seem to apply to a range of numerical techniques.

Until a few years ago, Casimir forces and other EM fluctuation-induced
interactions occupied an unusual position in this tableau. Realistic,
general numerical methods to solve for Casimir forces were simply
unavailable; solutions were limited to special high-symmetry geometries
(and often to special materials like perfect metals) that are amenable
to analytical and semi-analytical approaches. This is not to say that
there were not, \emph{in principle}, decades-old theoretical frameworks
capable of describing fluctuations for arbitrary geometries and materials,
but practical techniques for \emph{evaluating} these theoretical descriptions
on a computer have only been demonstrated in the last few years~\cite{Rodriguez07:PRL,Rodriguez07:PRA,Pasquali09,RodriguezMc09:PRA,McCauleyRo10:PRA,XiongCh09,XiongTo10:arxiv,emig01,maianeto05,emig06,Lambrecht06,Emig07,Rahi07,Kenneth08,MaiaNeto08,Reynaud08,ReidRo09,Rahi09:PRD,Lambrecht09}.
In almost all cases, these approaches turn out to be closely related
to computational methods from \emph{classical} EM, which is fortunate
because it means that Casimir computations can exploit decades of
progress in computational classical EM once the relationship between
the problems becomes clear. The long delay in developing numerical
methods for Casimir interactions, from the time the phenomenon was
first proposed in 1948~\cite{casimir}, can be explained by three
factors. First, accurate measurements of Casimir forces were first
reported only in 1997~\cite{Lamoreaux97} and experimental interest
in complex Casimir geometries and materials has only recently experienced
dramatic growth due to the progress in fabricating nanoscale mechanical
devices. Second, even the simplest numerical prediction of a single
force requires the equivalent of a large number of classical EM simulations,
a barrier to casual numerical experimentation. Third, there have historically
been many equivalent theoretical formulations of Casimir forces, but
some formulations are much more amenable to computational solution
than others, and these formulations are often couched in a language
that is opaque to researchers from classical computational EM.

This purpose of this review is to survey the available and proposed
numerical techniques for evaluating Casimir forces, energies, torques,
and related interactions, emphasizing their relationships to standard
classical-EM methods. Our goal is not to identify a {}``best'' method,
but rather to illuminate the strengths and weaknesses of each approach,
highlighting the conclusions that can be gleaned from the classical
experience. We will review an intellectual framework in which to evaluate
different numerical techniques, comparing them along several axes
for which quasi-independent choices of approach can be made. We will
also emphasize a few key departures of Casimir problems from ordinary
classical EM, such as the necessity of imaginary- or complex-frequency
solutions of Maxwell's equations and the need for wide-bandwidth analyses,
that impact the adaptation of off-the-shelf computational methods.

\section{Characterization of numerical methods: Three axes}

Numerical methods from distinct groups or research papers often differ
in several ways simultaneously, complicating the task of directly
comparing or even describing them. In order to organize one's understanding
of numerical approaches, it is useful to break them down along three
axes of comparison, representing (roughly) independent choices in
the design of a method:
\begin{itemize}
\item What \textbf{problem} does the method solve\textemdash{}even within
a single area such as classical EM, there are several conceptually
different \emph{questions} that one can ask and several \emph{ways
of asking} them that lead to different categories of methods.
\item What \textbf{basis} is used to express the unknowns\textemdash{}how
the \emph{infinite} number of unknowns in the exact partial differential
equation (PDE) or integral equation are reduced to a \emph{finite}
number of unknowns for solution on a computer.
\item What \textbf{solution technique} is used to determine these unknowns\textemdash{}even
with the same equations and the same unknowns, there are vast differences
among the types of direct, sparse, and iterative methods that can
be used to attack the problem, and the efficient application of a
particular solution technique to a particular problem is sometimes
a research task unto itself.
\end{itemize}
In this section, we briefly summarize the available problems, basis
choices, and solution techniques for Casimir problems. In subsequent
sections, we then discuss in more detail the specific approaches that
have currently been demonstrated or proposed.

\subsection{Posing Casimir problems}

In classical EM, there are several types of problems that are typically
posed~\cite[appendix D]{JoannopoulosJo08-book}, such as computing
source-free time-harmonic eigensolutions $\vec{E},\vec{H}\sim e^{-i\omega t}$
and eigenfrequencies $\omega$, computing time-harmonic fields resulting
from a time-harmonic current source $\vec{J}\sim e^{-i\omega t}$,
or computing the time-dependent fields created by an arbitrary time-dependent
source $\vec{J}(t)$ starting at $t=0$. Although these are all closely
mathematically related, and in some sense the solution of one problem
can give solutions to the other problems, they lead to very different
types of numerical simulations.

In a similar way, despite the fact that different formulations of
Casimir-interaction problems are ultimately mathematically equivalent
(although the equivalencies are often far from obvious)---and are
usually answering the same conceptual question, such as what is the
force or interaction energy for some geometry---each one leads most
naturally to distinct classes of computational methods. Here, we exclude
formulations such as proximity-force ({}``parallel-plate'') approximations~\cite{Derjaguin56,bordag01,Bordag06},
pairwise summation of Casimir--Polder forces (valid in the dilute-gas
limit)~\cite{Casimir48:polder,Tajmar04,Sedmik06}, and ray optics~\cite{Jaffe04,Hertzberg07,Jaffe05:approach,Zaheer07},
that are useful in special cases but represent uncontrolled approximations
if they are applied to arbitrary geometries. Although at some point
the distinctions are blurred by the mathematical equivalencies, we
can crudely categorize the approaches as:
\begin{itemize}
\item Computing the eigenfrequencies $\omega_{n}$ and summing the zero-point
energy $\sum_{n}\frac{\hbar\omega_{n}}{2}$~\cite{casimir,Enk95}.
See Sec.~\ref{sec:eigenmode-summation}.
\item Integrating the mean energy density or force density (stress tensor),
by evaluating field correlation functions $\langle E_{i}E_{j}\rangle_{\omega}$
and $\langle H_{i}H_{j}\rangle_{\omega}$ in terms of the classical
EM Green's functions at $\omega$ via the fluctuation--dissipation
theorem~\cite{Rodriguez07:PRL,Rodriguez07:PRA,Pasquali09,RodriguezMc09:PRA,McCauleyRo10:PRA,XiongCh09,XiongTo10:arxiv}.
See Sec.~\ref{sec:fluctuation-dissipation}.
\item Evaluating a path-integral expression for the interaction energy (or
its derivative), constrained by the boundary conditions---usually,
portions of the path integrals are performed analytically to express
the problem in terms of classical scattering matrices or Green's functions
at each~$\omega$~\cite{emig01,maianeto05,emig06,Lambrecht06,Emig07,Rahi07,Kenneth08,MaiaNeto08,Reynaud08,ReidRo09,Rahi09:PRD,Lambrecht09}.
See Sec.~\ref{sec:path-integrals}.
\end{itemize}
In each case, the result must be summed/integrated over all frequencies
$\omega$ to obtain the physical result (corresponding to thermodynamic/quantum
fluctuations at all frequencies). The relationship of the problem
to \emph{causal} Green's functions (fields appear \emph{after} currents)
means that the integrand is analytic for $\Im\omega\geq0$~\cite{Jackson98}.
As a consequence, there is a choice of \emph{contours} of $\omega$
integration in the upper-half complex plane, which is surprisingly
important---it turns out that the integrands are wildly oscillatory
on the real-$\omega$ axis and require accurate integration over a
huge bandwidth, whereas the integrands are much better-behaved along
the imaginary-$\omega$ axis ({}``Wick-rotated'' or {}``Matsubara''
frequencies). This means that Casimir calculations almost always involve
classical EM problems evaluated at \emph{complex or imaginary frequencies},
as is discussed further in Sec.~\ref{sec:complex-frequency}. The
nonzero-temperature case, where the integral over imaginary frequencies
becomes a sum (numerically equivalent to a trapezoidal-rule approximation),
is discussed in Sec.~\ref{sec:finite-temperature}.

There is also another way to categorize the problem to be solved:
whether one is solving a partial differential equation (\textbf{PDE})
or an \textbf{integral equation}. In a PDE, one has \emph{volumetric
unknowns}: fields or other functions at every point in space, related
to one another \emph{locally} by derivatives and so on. In an integral
equation, one typically has \emph{surface unknowns}: the fields or
currents on the \emph{boundaries} between piecewise-homogeneous regions,
related to one another \emph{non-locally} by the Green's functions
of the homogeneous regions (typically known analytically) \cite{chew01,Volakis01}
(described further in Sec.~\ref{sub:BEM-stress}). The key point
is to take advantage of the common situation in which one has piecewise-constant
materials, yielding a \emph{surface integral equation}. (There are
also \emph{volume integral equations} for inhomogeneous media~\cite{ShaubertWi84},
as well as hybrid integral/PDE approaches~\cite{chew01}, but these
are less common.) There are other hybrid approaches such as \emph{eigenmode
expansion}~\cite{Willems95,Bienstman01,BienstmanBa02}, also called
\emph{rigorous coupled-wave analysis} (RCWA)~\cite{MoharamGr95,MoharamPo95}
or a \emph{cross-section} method~\cite{Katsenelenbaum98}: a structure
is broken up along one direction into piecewise-constant cross-sections,
and the unknown fields at the \emph{interfaces} between cross-sections
are propagated in the uniform sections via the \emph{eigenmodes} of
those cross-sections (computed analytically or numerically by solving
the PDE in the cross-section). Eigenmode expansion is most advantageous
for geometries in which the cross-section is constant over substantial
regions, just as integral-equation methods are most advantageous to
exploit large homogeneous regions.

\subsection{Choices of basis\label{sub:Choices-of-basis}}

Casimir problems, for the most part, reduce to solving classical EM
linear PDEs or integral equations where the unknowns reside in an
infinite-dimensional vector space of functions. To \emph{discretize}
the problem approximately into a finite number $N$ of unknowns, these
unknown functions must be expanded in some finite \emph{basis }(that
converges to the exact solution as $N\to\infty$). There are three
typical types of basis:
\begin{itemize}
\item \textbf{Finite differences}~\cite{Taflove00,Christ87,strikwerda89}
(\textbf{FD}): approximate a function $f(x)$ by its values on some
uniform grid with spacing $\Delta x$, approximate derivatives by
some difference expression {[}e.g. second-order center differences
$f'(x)\approx\frac{f(x+\Delta x)-f(x-\Delta x)}{2\Delta x}+O(\Delta x^{2})${]}
and integrals by summations (e.g. a trapezoidal rule).
\item \textbf{Finite-element methods}~\cite{Volakis01,chew01,Jin02,Zhu06}
(\textbf{FEM}): divide space into geometric \emph{elements} (e.g.
triangles/tetrahedra), and expand an unknown $f(x)$ in a simple \emph{localized
}basis expansion for each element (typically, low-degree polynomials)
with some continuity constraints. (FD methods are viewable as special
cases of FEMs for uniform grids.) For an integral-equation approach,
where the unknowns are functions on surfaces, the same idea is typically
called a \textbf{boundary-element method} (\textbf{BEM})~\cite{Hackbush89,bonnet99,chew01,Jin02,Volakis01}.%
\footnote{The name \emph{method of moments} is also commonly applied to BEM
techniques for EM. However, this terminology is somewhat ambiguous,
and can refer more generally to Galerkin or other weighted-residual
methods (and historically referred to monomial test functions, yielding
statistical {}``moments'')~\cite{boyd01:book}.%
}
\item \textbf{Spectral} methods~\cite{boyd01:book}: expand functions in
a non-localized complete basis, truncated to a finite number of terms.
Most commonly, Fourier series or related expansions are used (cosine
series, Fourier--Bessel series, spherical or spheroidal harmonics,
Chebyshev polynomials, \emph{etc.}).
\end{itemize}
Finite differences have the advantage of simplicity of implementation
and analysis, and the disadvantages of uniform spatial resolution
and relatively low-order convergence (errors typically $\sim\Delta x^{2}$~\cite{Taflove00}
or even $\sim\Delta x$ in the presence of discontinuous materials
unless special techniques are used~\cite{Ditkowski01,OskooiKo09}).
FEMs can have nonuniform spatial resolution to resolve disparate feature
sizes in the same problem, at a price of much greater complexity of
implementation and solution techniques, and can have high-order convergence
at the price of using complicated curved elements and high-order basis
functions. Spectral methods can have very high-order or possibly exponential
({}``spectral'') convergence rates~\cite{boyd01:book} that can
even suit them to analytical solution---hence, spectral methods were
the dominant technique before the computer era and are typically the
first class of methods that appear in any field, such as in Mie's
classic solution of wave scattering from a sphere~\cite{Stratton41}.
However, exponential convergence is usually obtained only if all discontinuities
and singularities are taken explicitly into account in the basis~\cite{boyd01:book}.
With discontinuous materials, this is typically only practical for
very smooth, high-symmetry geometries like spheres, cylinders, and
so on; the use of a generic Fourier/spectral basis for arbitrary geometries
reduces to a brute-force method that is sometimes very convenient~\cite{Johnson2001:mpb},
but may have unremarkable convergence rates~\cite{boyd01:book,Johnson2001:mpb,KuoTi08}.
BEMs require the most complicated implementation techniques, because
any nontrivial change to the Green's functions of the homogeneous
regions (e.g. a change in dimensionality, boundary conditions, or
material types) involves tricky changes to the singular-integration
methods required to assemble the matrix~\cite{Taylor03,SladekSl98,TongCh06}
and to the fast-solver methods mentioned in Sec.~\ref{sub:Solution-techniques}.

Given FEM/BEM or spectral basis functions $b_{n}(x)$ and a linear
equation $\hat{A}u(x)=v(x)$ for an unknown function $u$ in terms
of a linear differential/integral operator $\hat{A}$, there are two
common ways~\cite{boyd01:book} to obtain a finite set of $N$ equations
to determine the $N$ unknown coefficients $c_{n}$ in $u(x)\approx\sum_{n}c_{n}b_{n}(x)$.
One is a \textbf{collocation }method: require that $(\hat{A}u-v)|_{x_{n}}=0$
be satisfied at \emph{N} collocation points $x_{n}$. The other is
a \textbf{Galerkin} method: require that $\langle b_{k},\hat{A}u-v\rangle=0$
be satisfied for $k=1,\ldots,N$, where $\langle\cdot,\cdot\rangle$
is some inner product on the function space. Both approaches result
in an $N\times N$ matrix equation of the form $A\vec{u}=\vec{v}$.
A Galerkin method has the useful property that if $\hat{A}$ is Hermitian
and/or definite then the matrix $A_{kn}=\langle b_{k},\hat{A}b_{n}\rangle$
has the same properties.

The specific situation of vector-valued unknowns in EM creates additional
considerations for the basis functions. In order to obtain center-difference
approximations for all the field components, FD methods for EM typically
use a staggered \textbf{Yee grid}~\cite{Taflove00,Christ87}, in
which each component of the EM fields is offset onto its own $\frac{\Delta x}{2}$-shifted
grid. In FEMs for EM, in order to maintain the appropriate continuity
conditions for curl or divergence operators, one uses special classes
of \emph{vector-valued} basis functions such as N�d�lec elements~\cite{Nedelec80,Jin02}.
In BEMs for EM, vector-valued \textbf{RWG }(Rao, Wilton, and Glisson)
basis functions~\cite{RWG82} (or generalizations thereof~\cite{CaiYu02})
are used in order to enforce a physical continuity condition on surface
currents (to preclude accumulation of charge at element edges); see
also Fig.~\ref{fig:crossed-capsules} in Sec.~\ref{sub:BEM-stress}.
A spectral integral-equation method for EM with cylindrical or spherical
scatterers is sometimes called a \textbf{multipole-expansion} method~\cite{Yasumoto05},
since the obvious spectral basis is equivalent to expanding the scattered
fields in terms of multipole moments.

\subsection{Solution techniques for linear equations\label{sub:Solution-techniques}}

Given a particular problem and basis choice, one at the end obtains
some $N\times N$ set of linear equations $A\vec{x}=\vec{b}$ to solve
(or possibly eigenequations $A\vec{x}=\lambda B\vec{x}$).%
\footnote{This applies equally well, if somewhat indirectly, to the path-integral
expressions of Sec.~\ref{sub:scattering-matrix} where one evaluates
a log determinant or a trace of an inverse, since this is done using
either eigenvalues or the same matrix factorizations that are used
to solve $A\vec{x}=\vec{b}$.%
} Note also that a \emph{single} Casimir-force calculation requires
the solution of \emph{many} such equations, at the very least for
an integral over frequencies (see Sec.~\ref{sec:complex-frequency}).
There are essentially three ways to solve such a set of equations:
\begin{itemize}
\item \textbf{Dense-direct} solvers: solve $A\vec{x}=\vec{b}$ using direct
matrix-factorization methods (e.g. Gaussian elimination),%
\footnote{Technically, all eigensolvers for $N>4$ are necessarily iterative,
but modern dense-eigensolver techniques employ direct factorizations
as steps of the process~\cite{Trefethen97}.%
} requiring $O(N^{2})$ storage and $O(N^{3})$ time~\cite{Trefethen97}.
\item \textbf{Sparse-direct} solvers~\cite{Davis06}: if $A$ is \emph{sparse}
(mostly zero entries), use similar direct matrix-factorization methods,
but cleverly re-arranged in an attempt to preserve the sparsity. Time
and storage depend strongly on the \emph{sparsity pattern} of $A$
(the pattern of nonzero entries).
\item \textbf{Iterative methods}~\cite{Trefethen97,barrett94,bai00}: repeatedly
improve a guess for the solution $\vec{x}$ (usually starting with
a random or zero guess), only referencing $A$ via repeated matrix--vector
multiplies. Time depends strongly on the properties of $A$ and the
iterative technique, but typically requires only $O(N)$ storage.
Exploits any fast way {[}ideally $O(N)$ or $O(N\log N$){]} to multiply
$A$ by any arbitrary vector.
\end{itemize}
If the number $N$ of degrees of freedom is small, i.e. if the basis
converges rapidly for a given geometry, dense-direct methods are simple,
quick, and headache-free (and have a standard state-of-the-art implementation
in the free LAPACK library~\cite{anderson99}). For example, $N=1000$
problems can be solved in under a second on any modern computer with
a few megabytes of memory. Up to $N\sim10^{4}$ is reasonably feasible,
but $N=10^{5}$ requires almost 100~GB of memory and days of computation
time without a large parallel computer. This makes dense-direct solvers
the method of choice in simple geometries with a rapidly converging
spectral basis, or with BEM integral-equation methods for basic shapes
that can be accurately described by a few thousand triangular panels,
but they rapidly become impractical for larger problems involving
many and/or complex objects (or for moderate-size PDE problems even
in two dimensions).

In PDE methods with a localized (FD or FEM) basis, the matrices $A$
have a special property: they are \emph{sparse} (mostly zero). The
locality of the operators in a typical PDE means that each grid point
or element directly interacts only with a bounded number of neighbors,
in which case $A$ has only $O(N)$ nonzero entries and can be stored
with $O(N)$ memory. The process of solving $A\vec{x}=\vec{b}$, e.g.
computing the LU factorization $A=LU$ by Gaussian elimination~\cite{Trefethen97},
unfortunately, ordinarily destroys this sparsity: the resulting $L$
and $U$ triangular matrices are generally not sparse. However, the
pattern of nonzero entries that arises from a PDE is not random, and
it turns out that clever re-orderings of the rows and columns during
factorization can partially preserve sparsity for typical patterns;
this insight leads to \emph{sparse-direct} solvers~\cite{Davis06},
available via many free-software packages implementing different sparsity-preserving
heuristics and other variations~\cite{bai00}. The sparsity pattern
of $A$ depends on the dimensionality of the problem, which determines
the number of neighbors a given element interacts with. For meshes/grids
having nearest-neighbor interactions, a sparse-direct solver typically
requires $O(N)$ time and storage in 1d (where the matrices are band-diagaonal),
$O(N^{3/2})$ time with $O(N\log N)$ storage in 2d, and $O(N^{2})$
time with $O(N^{4/3})$ storage in 3d~\cite{DuffEr76,Davis06}. The
practical upshot is that sparse-direct methods work well for 1d and
2d PDEs, but can grow to be impractical in 3d. For BEM and spectral
methods, the interactions are not localized and the matrices are not
sparse, so sparse-direct methods are not directly applicable (but
see below for an indirect technique).

For the largest-scale problems, or for problems lacking a sparse $A$,
the remaining possibility is an \emph{iterative} method. In these
methods, one need only supply a fast way to multiply $A$ by an arbitrary
vector $y$, and the trick is to use this $Ay$ operation on a clever
sequence of vectors in such a way as to make an arbitrary initial
guess $x_{0}$ converge as rapidly as possible to the solution $x$,
ideally using only $O(N)$ storage. Many such techniques have been
developed~\cite{Trefethen97,barrett94,bai00}. The most favorable
situation for $Ax=b$ occurs when $A$ is Hermitian positive-definite,
in which case an ideal Krylov method called the \emph{conjugate-gradient}
method can be applied, with excellent guaranteed convergence properties~\cite{Trefethen97,barrett94},
and fortunately this is precisely the case that usually arises for
the imaginary-frequency Casimir methods below. There are two wrinkles
that require special attention, however. First, one must have a fast
way to compute $Ay$. If $A$ is sparse (as for PDE and FD methods),
then only $O(N)$ nonzero entries of $A$ need be stored (as above)
and $Ay$ can be computed in $O(N)$ operations. In a spectral method,
$A$ is generally dense, but for spectral PDE methods there are often
fast $O(N\log N)$ techniques to compute $Ay$ using only $O(N)$
storage ($A$ is stored implicitly), based on fast Fourier transform
(FFT) algorithms~\cite{boyd01:book,Johnson2001:mpb}. In a BEM, where
$A$ is again dense, a variety of sophisticated methods that require
only $O(N\log N)$ computation time and $O(N)$ storage to compute
$Ay$ (again storing $A$ implicitly) have been developed~\cite{PhillipsWh97,chew01,Volakis01,Jin02},
beginning with the pioneering fast-multipole method~(FMM)~\cite{GreengardRo85}.
These fast BEMs exploit the localized basis and the decaying, convolutional
nature of the Green's function to approximate long-range interactions
(to any desired accuracy). FMMs can be viewed as an approximate factorizations
into sparse matrices, at which point sparse-direct methods are also
applicable~\cite{GreengardGu09}. A second wrinkle is that the convergence
rates of iterative methods depend on the condition number of $A$
(the ratio of largest and smallest singular values)~\cite{Trefethen97,barrett94},
and condition numbers generally worsen as the ratio of the largest
and smallest lengthscales in the problem increases. To combat this,
users of iterative methods employ \textbf{preconditioning} techniques:
instead of solving $Ax=b$, one solves $KAx=Kb$ or similar, where
the preconditioning matrix $K$ is some crude \emph{approximate inverse}
for $A$ (but much simpler to compute than $A^{-1}$!) such that the
condition number of $KA$ is reduced~\cite{barrett94}. The difficulty
with this approach is that good preconditioners tend to be highly
problem-dependent, although a variety of useful approaches such as
incomplete factorization and coarse-grid/multigrid approximations
have been identified~\cite{barrett94,Trefethen97}. The upshot is
that, while the largest-scale solvers almost invariably use iterative
techniques, for any given class of physical problems it sometimes
takes significant research before the iterative approach becomes well-optimized.

\section{The impracticality of eigenmode summations\label{sec:eigenmode-summation}}

Perhaps the simplest way to express the Casimir energy, at zero temperature,
is as a sum of zero-point energies of all oscillating EM modes in
the system: \begin{equation}
U=\sum_{\omega}\frac{\hbar\omega}{2},\label{eq:U-sum}\end{equation}
where $\omega$ is the frequency of the mode ($\sim e^{-i\omega t}$)~\cite{casimir,milonni}.
That is, when the electromagnetic field is quantized into photons
with energy $\hbar\omega$, it turns out that the vacuum state in
the absence of photons is not empty, but rather has the energy equivalent
of {}``half a photon'' in each mode. The computational strategy
is then straightforward, in principle: compute the EM eigenfrequencies
$\omega$ in the problem by some numerical method (many techniques
are available for computing eigenfrequencies~\cite{Johnson2001:mpb,chew01})
and sum them to obtain $U$. Forces are then given by the derivative
of $U$ with respect to changes in the geometry, which could be approximated
by finite differences or differentiated analytically with a Hellman--Feynman
technique~\cite{Tannoudji77} (more generally, derivatives of any
computed quantity can be computed efficiently by an adjoint method~\cite{Strang07-CSE}).

Of course, $U$ in eq.~(\ref{eq:U-sum}) has the disadvantage of
being formally infinite, but this is actually a minor problem in practice:
as soon as one discretizes the problem into a finite number of degrees
of freedom (e.g., a finite number of grid points), the number of eigenfrequencies
becomes finite (with the upper bound representing a Nyquist-like frequency
of the grid). This is the numerical analogue~\cite{Rodriguez07:PRA}
of analytical regularization techniques that are applied to truncate
the same sum in analytical computations~\cite{casimir}. 
(These regularizations
do not affect energy \emph{differences} or forces for rigid-body motions.) 
Matters
are also somewhat subtle for dissipative or open systems~\cite{Lamoreaux05}.
But the most serious problem is that, even in the lossless case, this
sum is badly behaved: even when one differentiates with separation
$a$ to obtain a finite force $F=-\frac{\hbar}{2}\sum\frac{d\omega}{da}$,
the summand is wildly oscillatory and includes substantial contributions
from essentially \emph{every} frequency, which mostly cancel to leave
a tiny result~\cite{Ford93,Rodriguez07:PRA}. Numerically, therefore,
one must ostensibly compute \emph{all} of the modes, to high precision,
which requires $O(N^{3})$ time and $O(N^{2})$ storage (for a dense-direct
eigensolver~\cite{Trefethen97}) given $N$ degrees of freedom. This
is possible in simple 1d problems~\cite{Enk95,Rodriguez07:PRA},
but is impractical as a general approach.

Because of the mathematical equivalence of the different approaches
to the Casimir problem, the mode-summation method is sometimes useful
as a starting point to derive alternative formulations, but the end
result is invariably quite different in spirit from computing the
eigenfrequencies one by one and summing them. For example, if one
has a function $z(\omega)$ whose roots are the eigenfrequencies,
then one can equivalently write $U$, via the residue theorem of complex
analysis, as $U=\frac{1}{2\pi i}\oint_{C}\frac{\hbar\omega}{2}\frac{d[\ln z(\omega)]}{d\omega}d\omega$,
where $C$ is any closed contour in the complex-$\omega$ plane that
encloses the roots~\cite{Nesterenko98}. However, finding functions
whose roots are the eigenfrequencies naturally points towards Green's
functions (to relate different boundary conditions), and the contour
choices typically involve Wick rotation as in Sec.~\ref{sec:complex-frequency},
so this approach leads directly to imaginary-frequency scattering-matrix
techniques as in Sec.~\ref{sec:path-integrals}~\cite{Lambrecht09}.
A similar contour integral arises from a zeta-function regularization
of~(\ref{eq:U-sum})~\cite{Cognola01}.

\section{The complex-frequency plane and contour choices\label{sec:complex-frequency}}

In order to better understand the frequency integration/summation
in Casimir problems, it is illustrative to examine the analytical
formula for the simple case of two perfect-metal plates in vacuum
separated by a distance $a$, in which case it can be derived in a
variety of ways that the attractive force $F$ is given by~\cite{Lifshitz80}:\begin{equation}
\begin{split}F & =\frac{\hbar}{\pi^{2}c^{3}}\Re\left[\int_{0}^{\infty}d\omega\int_{1}^{\infty}dp\frac{p^{2}\omega^{3}}{e^{2ip(\omega+i0^{+})a/c}-1}\right]\\
 & =\Re\left[\int_{0}^{\infty}f(\omega)d\omega\right]=\Im\left[\int_{0}^{\infty}f(i\xi)d\xi\right]=\frac{\hbar c}{240a^{4}},\end{split}
\label{eq:Lifshitz-metal}\end{equation}
where $f(\omega)$ is the contribution of each frequency $\omega$
to the force and $p$ is related to the plate-parallel momentum of
the contributing modes/fluctuations. In this special case, the entire
integral can be performed analytically, but for parallel plates of
some finite permittivity $\varepsilon$ the generalization (the\emph{
Lifshitz formula}~\cite{Lifshitz80}) must be integrated numerically.
In practice, however, the formula and its generalizations are never
integrated in the form at left---instead, one uses the technique of
contour integration from complex analysis to \emph{Wick rotate} the
integral to imaginary frequencies $\omega=i\xi$, integrating over
$\xi$. (In fact, the formula is typically derived \emph{starting}
in imaginary frequencies, via a Matsubara approach~\cite{Lifshitz80}.)
In this section, we review why a trick of this sort is both \emph{possible}
and \emph{essential} in numerical computations for all of the methods
described below.

Wick rotation is always \emph{possible} as a consequence of causality.
It turns out that the frequency contributions $f(\omega)$ for arbitrary
materials and geometries, for all of the different formulations of
the Casimir force below, are ultimately expressed in terms of classical
EM \emph{Green's functions} at $\omega$: the EM fields in response
to time-harmonic currents $\vec{J}\sim e^{-i\omega t}$. As a consequence
of the causality of Maxwell's equations and physical materials---EM
fields always arise \emph{after} the source currents, not before---it
mathematically follows that the Green's functions must be analytic
functions (no poles or other singularities) when $\Im\omega>0$ (the
upper-half complex plane)~\cite{Jackson98}. Poles in the Green's
function correspond to eigenfrequencies or resonances of the source-free
Maxwell's equations, and must lie at $\Im\omega \leq 0$ for any physical
system with dissipative materials (with the poles approaching $\Im\omega=0^{-}$
in the idealized lossless limit). {[}One can easily see explicitly
that this is true for the $f(\omega)$ above: the poles result from
a vanishing denominator in the $p$ integrand, which only occurs for
purely real $\omega$ corresponding to the real-frequency modes trapped
between two perfect-metal plates.{]} As an elementary consequence
of complex analysis, this analyticity means that the $\int d\omega$
can be arbitrarily deformed to any contour in the upper-half complex-$\omega$
plane without changing the integration result.

\begin{figure}[t]
\begin{centering}
\includegraphics[width=1\columnwidth]{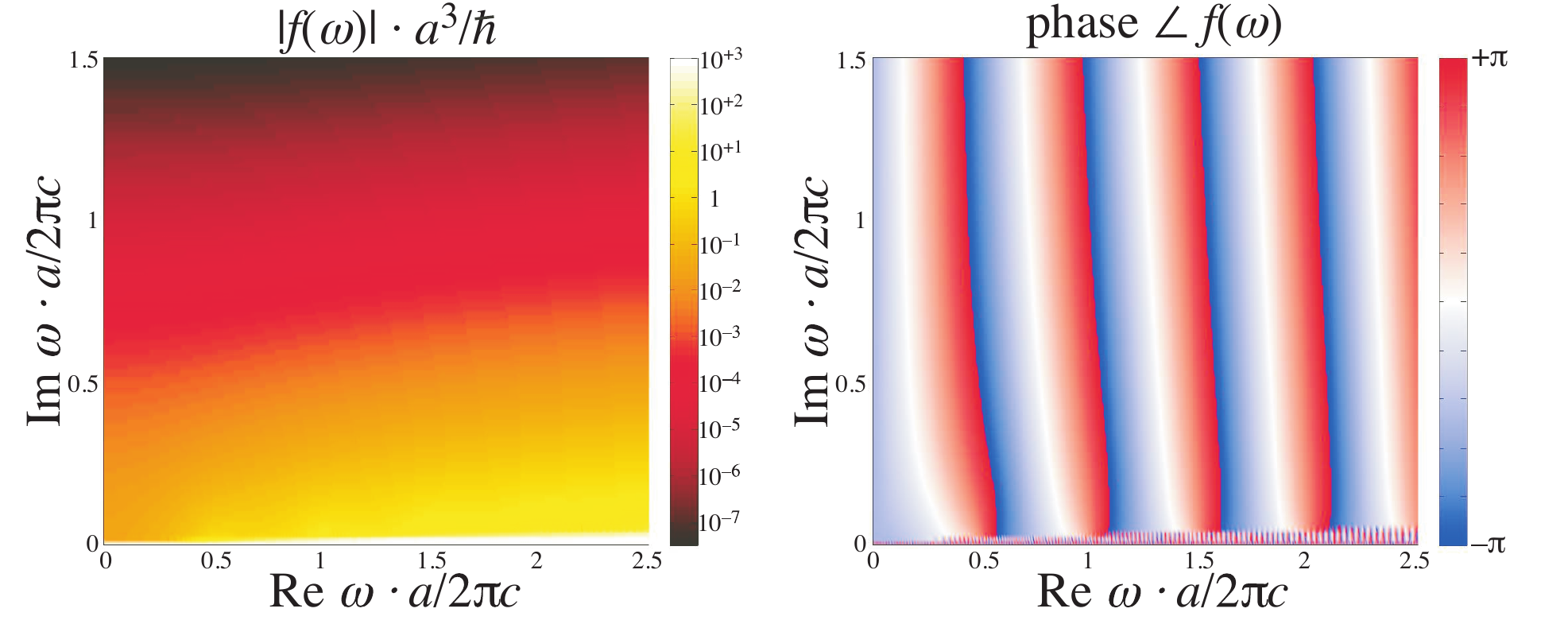}
\par\end{centering}
\caption{\label{fig:lifshitz-omega}Contributions $f(\omega)$ to the Casimir
force, from each fluctuation/mode frequency $\omega$, for two perfect-metal
plates with separation $a$, in the complex-$\omega$ plane. \emph{Left}:
magnitude $|f(\omega)|$. \emph{Right:} phase $\angle f(\omega)$.
{[}The magnitude is truncated at $10^{3}\hbar/a^{3}$, as it diverges
towards the real-$\omega$ axis, and some numerical artifacts (rapid
oscillations) are visible near the real-$\omega$ axis in the phase
due to difficulty in evaluating $f(\omega)$. {]} The key point is
that $f(\omega)$ is badly behaved (oscillatory and non-decaying)
along contours parallel to the real-$\omega$ axis, whereas $f(\omega)$
is nicely behaved (non-oscillatory and exponentially decaying) along
contours parallel to the imaginary-$\omega$ axis.}
\end{figure}

Wick rotation is \emph{essential} for computation because the frequency
contributions $f(\omega)$ to the force (or interaction energy or
other related quantities) are extremely ill-behaved near to the real-$\omega$
axis: they are wildly oscillatory and slowly decaying. For example,
the magnitude and phase of the function $f(\omega)$ are plotted in
the complex $\omega$ plane in Fig.~\ref{fig:lifshitz-omega}, where
the $p$ integral was evaluated numerically with a high-order Clenshaw--Curtis
quadrature scheme~\cite{Boyd87}.
 Merely evaluating $f(\omega)$ along the real-$\omega$ axis is difficult
because of singularities (which ultimately reduce the integral to
a summation over eigenfrequency-contributions as in Sec.~\ref{sec:eigenmode-summation});
in physical materials with dissipation, the real-$\omega$ axis is
non-singular but is still badly behaved because of poles (lossy modes)
located just below the axis. Along any contour parallel to the real-$\omega$
axis, the integrand is oscillatory (as can be seen from the phase
plot) and non-decaying (as can be seen from the magnitude plot): formally,
just as with the infinite summation over eigenmodes in Sec.~\ref{sec:eigenmode-summation},
one must integrate over an infinite bandwidth, regularized in some
way (e.g. by the Nyquist frequency placing an upper bound on $\omega$
for a finite grid), where the oscillations almost entirely cancel
to leave a tiny remainder (the force). (Any physical materials must
cease to polarize as $\omega\to\infty$ where the susceptibility vanishes~\cite{Jackson98},
which will make the force contributions eventually vanish as $\omega\to\infty$
even in~1d, but a very wide-bandwidth oscillatory integral is still
required.) This is a disaster for any numerical method---even when
one is only integrating an analytical expression such as the Lifshitz
formula, mere roundoff errors are a severe difficulty for real $\omega$.
Along the imaginary-$\omega$ axis, on the other hand (or any sufficiently
vertical contour), $f(\omega)$ is exponentially decaying and mostly
non-oscillatory---an ideal situation for numerical integration.

Therefore, in order for classical EM solvers to be used for Casimir
problems, they need to be adapted to solve Maxwell's equations at
complex or imaginary $\omega$. Although this sounds strange at first,
the frequency-domain problem actually becomes numerically \emph{easier}
in every way at imaginary $\omega$; this is discussed in more detail
in Sec.~\ref{sub:imaginary-Green}. In fact, one can even identify
an exact mathematical equivalence between a particular complex-$\omega$
contour and a \emph{real}-frequency system where an \emph{artificial
dissipation} has been introduced, as discussed in Sec.~\ref{sub:FDTD}
below---using this trick, one can actually use classical EM solvers
with no modification at all, as long as they handle dissipative media.
In any case, one needs an integral over frequencies to compute a physically
meaningful quantity, which means that solvers and material models,
not to mention any physical intuition used for guidance, must be valid
for more than just a narrow real-$\omega$ bandwidth (unlike most
problems in classical EM).

Numerically, it should be pointed out that the $f(i\xi)$ integrand
is smooth and exponentially decaying, and so the $\xi$ integral can
be approximated to high accuracy by an exponentially convergent quadrature
(numerical integration) scheme using evaluations at relatively few
points $\xi$. For example, one can use Gauss--Laguerre quadrature~\cite{StroudSe66},
Gaussian quadrature with an appropriate change of variables~\cite{piessens83},
or Clenshaw--Curtis quadrature with an appropriate change of variables~\cite{Boyd87}.

\section{Mean energy/force densities and the fluctuation--dissipation theorem\label{sec:fluctuation-dissipation}}

Another, equivalent, viewpoint on Casimir interactions is that they
arise from geometry-dependent fluctuations of the electromagnetic
fields $\vec{E}$ and $\vec{H}$, which on average have some nonzero
energy density and exert a force. If we can compute these average
fields, we can integrate the resulting energy density, stress tensors,
and so on, to obtain the energy, force, or other quantities of interest.
The good news is that there is a simple expression for those fluctuations
in terms of the \textbf{fluctuation--dissipation} theorem of statistical
physics: the correlation function of the fields is related to the
corresponding \emph{classical }Green's function~\cite{Lifshitz80}.
Ultimately, this means that any standard classical EM technique to
compute Green's functions (fields from currents) can be applied to
compute Casimir forces, with the caveat that the techniques must be
slightly modified to work at imaginary or complex frequencies as described
below.

\subsection{Background}

The temperature-$T$ correlation function for the fluctuating electric
field at a given frequency $\omega$ is given by~\cite{Lifshitz80}:
\begin{equation}
\langle E_{j}(\vec{x})E_{k}(\vec{x}')\rangle_{\omega}=-\frac{\hbar}{\pi}\Im\left[\omega^{2}G_{jk}^{E}(\omega;\vec{x},\vec{x}')\right]\coth(\hbar\omega/2k_{\mbox{B}}T),\label{eq:Ecorr-real}\end{equation}
where $G_{jk}^{E}=(\vec{G}_{k}^{E})_{j}$ is the classical dyadic
{}``photon'' Green's function, proportional%
\footnote{The electric field $\vec{E}(\vec{x})$ from a dipole current $\vec{J}=\delta^{3}(\vec{x}-\vec{x}')\hat{e}_{k}e^{-i\omega t}$
is $\vec{E}(\vec{x})=i\omega\vec{G}_{k}^{E}(\omega,\vec{x},\vec{x}')e^{-i\omega t}$. %
} to the relationship between an electric-dipole current in the $k$
direction at $\vec{x}'$ to the electric field at $\vec{x}$, and
solves \begin{equation}
\left[\nabla\times\mu(\omega,\vec{x})^{-1}\nabla\times{}-\omega^{2}\varepsilon(\omega,\vec{x})\right]\vec{G}_{k}^{E}(\omega,\vec{x},\vec{x}')=\delta^{3}(\vec{x}-\vec{x}')\hat{e}_{k},\label{eq:Green-real}\end{equation}
where $\varepsilon$ is the electric permittivity tensor, $\mu$ is
the magnetic permeability tensor, and $\hat{e}_{k}$ is a unit vector
in direction $k$. Similarly, the magnetic-field correlation function
is \begin{equation}
\langle H_{j}(\vec{x})H_{k}(\vec{x}')\rangle_{\omega}=-\frac{\hbar}{\pi}\Im\left[\omega^{2}G_{jk}^{H}(\omega;\vec{x},\vec{x}')\right]\coth(\hbar\omega/2k_{\mbox{B}}T).\label{eq:Hcorr-real}\end{equation}
The magnetic Green's function $\mathbf{G}^{H}$ can be defined in
two essentially equivalent ways. The first is as derivatives $\frac{1}{\omega^{2}\mu(\vec{x})}\nabla\times\mathbf{G}^{E}\times\nabla'\frac{1}{\mu(\vec{x}')}$
of the electric Green's function $G_{jk}^{E}(\vec{x},\vec{x}')$,
where $\nabla$ and $\nabla'$ denote derivatives with respect to
$\vec{x}$ and $\vec{x}'$ ($\nabla'$ acting to the left), respectively~\cite{Lifshitz80}.
The second way to define $\mathbf{G}^{H}$ is proportional to the
magnetic field in response to a magnetic-dipole current, analogous
to (\ref{eq:Green-real}):\begin{equation}
\left[\nabla\times\varepsilon(\omega,\vec{x})^{-1}\nabla\times{}-\omega^{2}\mu(\omega,\vec{x})\right]\vec{G}_{k}^{H}(\omega,\vec{x},\vec{x}')=\delta^{3}(\vec{x}-\vec{x}')\hat{e}_{k},\label{eq:Green-real-H}\end{equation}
which can be more convenient for numerical calculation~\cite{RodriguezMc09:PRA}.
These two definitions are related~\cite{Buhmann09} by $\mathbf{G}^{H}=\frac{1}{\omega^{2}\mu(\vec{x})}\nabla\times\mathbf{G}^{E}\times\nabla'\frac{1}{\mu(\vec{x}')}-\frac{1}{\omega^{2}\mu(\vec{x}')}\delta(\vec{x}-\vec{x}')I$
(with $I$ being the $3\times3$ identity matrix),%
\footnote{This can be seen more explicity by substituting $\mathbf{G}^{H}=\frac{1}{\omega^{2}}\frac{1}{\mu}\nabla\times\mathbf{G}^{E}\times\nabla'\frac{1}{\mu'}-\frac{1}{\omega^{2}\mu'}\delta$
into (\ref{eq:Green-real-H}), with $\delta$ denoting $\delta(\vec{x}-\vec{x}')I$
and $\mu$ or $\mu'$ denoting $\mu(\vec{x})$ or $\mu(\vec{x}')$,
respectively. In particular, $[\nabla\times\frac{1}{\varepsilon}\nabla\times{}-\omega^{2}\mu](\frac{1}{\omega^{2}}\frac{1}{\mu}\nabla\times\mathbf{G}^{E}\times\nabla'\frac{1}{\mu'}-\frac{1}{\omega^{2}\mu'}\delta)$
yields $\nabla\times[\frac{1}{\omega^{2}\varepsilon}\nabla\times\frac{1}{\mu}\nabla\times\mathbf{G}^{E}-\mathbf{G}^{E}]\times\nabla'\frac{1}{\mu'}-\nabla\times\frac{1}{\omega^{2}\mu'\varepsilon}\nabla\times\delta+\delta$,
which via (\ref{eq:Green-real}) gives $+\nabla\times\frac{1}{\omega^{2}\varepsilon}\delta\times\nabla'\frac{1}{\mu'}-\nabla\times\frac{1}{\omega^{2}\mu'\varepsilon}\nabla\times\delta+\delta=\delta$
as desired, where in the last step we have used the fact that $\delta\times\nabla'=\nabla\times\delta$
{[}since $\nabla\times{}$ is antisymmetric under transposition and
$\nabla'\delta(\vec{x}-\vec{x}')=-\nabla\delta(\vec{x}-\vec{x}')${]}.%
} where the second (diagonal) term has no effect on energy differences
or forces below and is therefore irrelevant. Now, these equations
are rather nasty along the real-$\omega$ axis: not only will there
be poles in $\mathbf{G}$ just below the axis corresponding to lossy
modes, but in the limit where the dissipative losses vanish ($\varepsilon$
and $\mu$ become real), the combination of the poles approaching
the real axis with the $\Im$ in the correlation function results
in a delta function at each pole%
\footnote{This follows from the standard identity that the limit $\Im[(x+i0^{+})^{-1}]$,
viewed as a distribution, yields $-\pi\delta(x)$~\cite{Gelfand64}.%
} and integrals of the correlation functions turn into sums over modes
as in Sec.~\ref{sec:eigenmode-summation}. However, the saving grace,
as pointed out in Sec.~\ref{sec:complex-frequency}, is that Green's
functions are \emph{causal}, allowing us to transform any integral
over all real fluctuation frequencies into an integral over \emph{imaginary}
fluctuation frequencies $\omega=i\xi$. The $\coth$ factor has poles
that alter this picture, but we will eliminate those for now by considering
only the $T=0^+$ case where $\coth(+\infty)=1$, returning to nonzero
temperatures in Sec.~\ref{sec:finite-temperature}.

\subsubsection{Energy density}

In particular, to compute the Casimir energy $U$, we merely integrate
the classical energy density in the EM field~\cite{Jackson98} over
all positions and all fluctuation frequencies, Wick-rotated to an
integral over imaginary frequencies, resulting in the expression:
\begin{equation}
U=\int_{0}^{\infty}d\xi\int\frac{1}{2}\left[\frac{d(\xi\varepsilon)}{d\xi}\langle|\vec{E}|^{2}\rangle_{i\xi}+\frac{d(\xi\mu)}{d\xi}\langle|\vec{H}|^{2}\rangle_{i\xi}\right]d^{3}\vec{x},\label{eq:U-mean-EH}\end{equation}
where we have simplified to the case of isotropic materials (scalar
$\varepsilon$ and $\mu$). At thermodynamic equilibrium, this expression
remains valid even for arbitrary dissipative/dispersive media thanks
to a direct equivalence with a path-integral expression~\cite{MiltonWa10:arxiv},
which is not obvious from the classical viewpoint in which the energy
density is usually only derived in the approximation of negligible
absorption~\cite{Jackson98}. (Thanks to the relationship between
the Green's function and the local density of states~\cite{Economou06},
there is also a direct equivalence between this energy integral and
eigenmode summation~\cite{Rodriguez07:PRA}.) In the common case
where $\mu$ has negligible frequency dependence (magnetic responses
are usually negligible at the short wavelengths where Casimir forces
are important, so that $\mu\approx\mu_{0}$), we can use the identity%
\footnote{Lest the application of this field identity appear too glib, we can
also obtain the same equality directly from the Green's functions
in the correlation functions. We have $\int\mu\langle|\vec{H}|^{2}\rangle=\frac{\hbar}{\pi}\tr\int\xi^{2}\mu\mathbf{G}^{H}(\vec{x},\vec{x})$,
and from the identity after (\ref{eq:Green-real-H}) we know that
$\xi^{2}\mu\mathbf{G}^{H}=-\nabla\times\mathbf{G}\times\nabla'\frac{1}{\mu'}+\delta$.
However, because $\nabla\times{}$ is self-adjoint~\cite{JoannopoulosJo08-book},
we can integrate by parts to move $\nabla\times{}$ from the first
argument/index of $\mathbf{G}^{E}$ to the second, obtaining $-\mathbf{G}^{E}\times\nabla'\frac{1}{\mu'}\times\nabla'=\xi^{2}\varepsilon'\mathbf{G}^{E}-\delta$
from the first term under the integral. (Here, we employ the fact
that $\mathbf{G}^{E}$ is real-symmetric at imaginary $\omega=i\xi$,
from Sec.~\ref{sub:imaginary-Green}, to apply (\ref{eq:Green-imag})
to the second index/argument instead of the first.) This cancels the
other delta from $\xi^{2}\mu\mathbf{G}^{H}$ and leaves $\xi^{2}\varepsilon\mathbf{G}^{E}$,
giving $\varepsilon\langle|\vec{E}|^{2}\rangle$ as desired.%
} that $\int\varepsilon|\vec{E}|^{2}=\int\mu|\vec{H}|^{2}$ for fields
at any given frequency~\cite{JoannopoulosJo08-book} to simplify
this expression to~\cite{Rodriguez07:PRA}: \begin{equation}
U=\int_{0}^{\infty}d\xi\int\frac{1}{2\xi}\frac{d(\xi^{2}\varepsilon)}{d\xi}\langle|\vec{E}|^{2}\rangle_{i\xi}d^{3}\vec{x}.\label{eq:U-mean-E}\end{equation}
Here, the zero-temperature imaginary-frequency mean-square electric
field is given by: \begin{equation}
\langle|\vec{E}(\vec{x})|^{2}\rangle_{i\xi}=\frac{\hbar}{\pi}\xi^{2}\tr\mathbf{G}^{E}(i\xi;\vec{x},\vec{x}),\label{eq:E2-imag}\end{equation}
where $\tr$ denotes the trace $\sum_{j}G_{jj}^{E}$ and the $\Im$
has disappeared compared to (\ref{eq:Ecorr-real}) because $\mathbf{G}^{E}(i\xi$)
is real and the $\Im$ cancels the $i$ in $d\omega\to i\, d\xi$.

Equation~(\ref{eq:EE-imag}) may at first strike one as odd, because
one is evaluating the Green's function (field) at $\vec{x}$ from
a source at $\vec{x}$, which is formally infinite. This is yet another
instance of the formal infinities that appear in Casimir problems,
similar to the infinite sum over modes in Sec.~\ref{sec:eigenmode-summation}.
In practice, this is not a problem either analytically or numerically.
Analytically, one typically regularizes the problem by subtracting
off the vacuum Green's function (equivalent to only looking at the
\emph{portion} of the fields at $\vec{x}$ which are reflected off
of inhomogeneities in $\varepsilon$ or $\mu$)~\cite{Lifshitz80}.
Numerically, in an FD or FEM method with a finite grid, the Green's
function is everywhere finite (the grid is its own regularization)~\cite{Rodriguez07:PRA}.
In a BEM, the Green's function is explicitly written as a sum of the
vacuum field and scattered fields, so the former can again be subtracted
analytically~\cite{Rodriguez07:PRA}. As in Sec.~\ref{sec:eigenmode-summation},
these regularizations do not affect physically observable quantities
such as forces or energy differences, assuming rigid-body motion.

\subsubsection{The remarkable imaginary-frequency Green's function\label{sub:imaginary-Green}}

This imaginary-frequency Green's function is actually a remarkably
nice object. Wick-rotating eq.~(\ref{eq:Green-real}), it satisfies:
\begin{equation}
\left[\nabla\times\mu(i\xi,\vec{x})^{-1}\nabla\times{}+\xi^{2}\varepsilon(i\xi,\vec{x})\right]\vec{G}_{k}^{E}(i\xi,\vec{x},\vec{x}')=\delta^{3}(\vec{x}-\vec{x}')\hat{e}_{k}.\label{eq:Green-imag}\end{equation}
Because of causality, it turns out that $\varepsilon$ and $\mu$
are strictly real-symmetric and positive-definite (in the absence
of gain) along the imaginary-frequency axis, even for dissipative/dispersive
materials~\cite{Jackson98}. Furthermore, the operator $\nabla\times\mu^{-1}\nabla\times{}$
is real-symmetric positive-semidefinite for a positive-definite real-symmetric
$\mu$~\cite{JoannopoulosJo08-book}. Thus, the entire bracketed
operator $[\cdots]$ in eq.~(\ref{eq:Green-imag}) is \emph{real-symmetric
positive-definite} for $\xi>0$, which lends itself to some of the
best numerical solution techniques (Cholesky decomposition~\cite{Trefethen97},
tridiagonal QR~\cite{Trefethen97}, conjugate gradients~\cite{Trefethen97,barrett94},
and Rayleigh-quotient methods~\cite{bai00}). (This definiteness
is also another way of seeing the lack of poles or oscillations for
$\omega=i\xi$.) It follows that the integral operator whose kernel
is $\mathbf{G}^{E}$, i.e. the inverse of the $[\cdots]$ operator
in eq.~(\ref{eq:Green-imag}), is also real-symmetric positive-definite,
which is equally useful for integral-equation methods.

In vacuum, the 3d real-$\omega$ Green's function $\sim e^{i\omega|\vec{x}-\vec{x}'|/c}/|\vec{x}-\vec{x}'|$~\cite{Jackson98}
is Wick-rotated to $\sim e^{-\xi|\vec{x}-\vec{x}'|/c}/|\vec{x}-\vec{x}'|$,
an \emph{exponentially decaying, non-oscillatory} function. This is
yet another way of understanding why, for $\omega=i\xi$, there are
no interference effects and hence no {}``modes'' (poles in $\mathbf{G}$),
and integrands tend to be non-oscillatory and exponentially decaying
(as $\xi\to\infty$, $\mathbf{G}$ becomes exponentially short-ranged
and does not {}``see'' the interacting objects, cutting off the
force contributions). (It also means, unfortunately, that a lot of
the most interesting phenomena in classical EM, which stem from interference
effects and resonances, may have very limited consequences for Casimir
interactions.)

One other property we should mention is that the operator becomes
semidefinite for $\xi=0$, with a nullspace encompassing any static
field distribution ($\nabla\phi$ for any scalar $\phi$). This corresponds
to the well-known singularity of Maxwell's equations at zero frequency~\cite{ZhaoCh00,EpsteinGr09},
where the electric and magnetic fields decouple~\cite{Jackson98}.
Since we eventually integrate over $\xi$, the measure-zero contribution
from $\xi=0$ does not actually matter, and one can use a quadrature
scheme that avoids evaluating $\xi=0$. However, in the nonzero-temperature
case of Sec.~\ref{sec:finite-temperature} one obtains a sum over
discrete-$\xi$ contributions, in which case the zero-frequency term
is explicitly present. In this case, $\xi=0$ can be interpreted if
necessary as the limit $\xi\to0^{+}$ (which can be obtained accurately
in several ways, e.g. by Richardson extrapolation~\cite{Press92},
although some solvers need special care to be accurate at low frequency~\cite{ZhaoCh00,EpsteinGr09});
note, however, that there has been some controversy about the zero-frequency
contribution in the unphysical limit of perfect/dissipationless metals~\cite{Hoye05}.

\subsubsection{Stress tensor\label{sub:Stress-tensor}}

\begin{figure}[t]

\begin{centering}
\includegraphics[width=0.5\columnwidth]{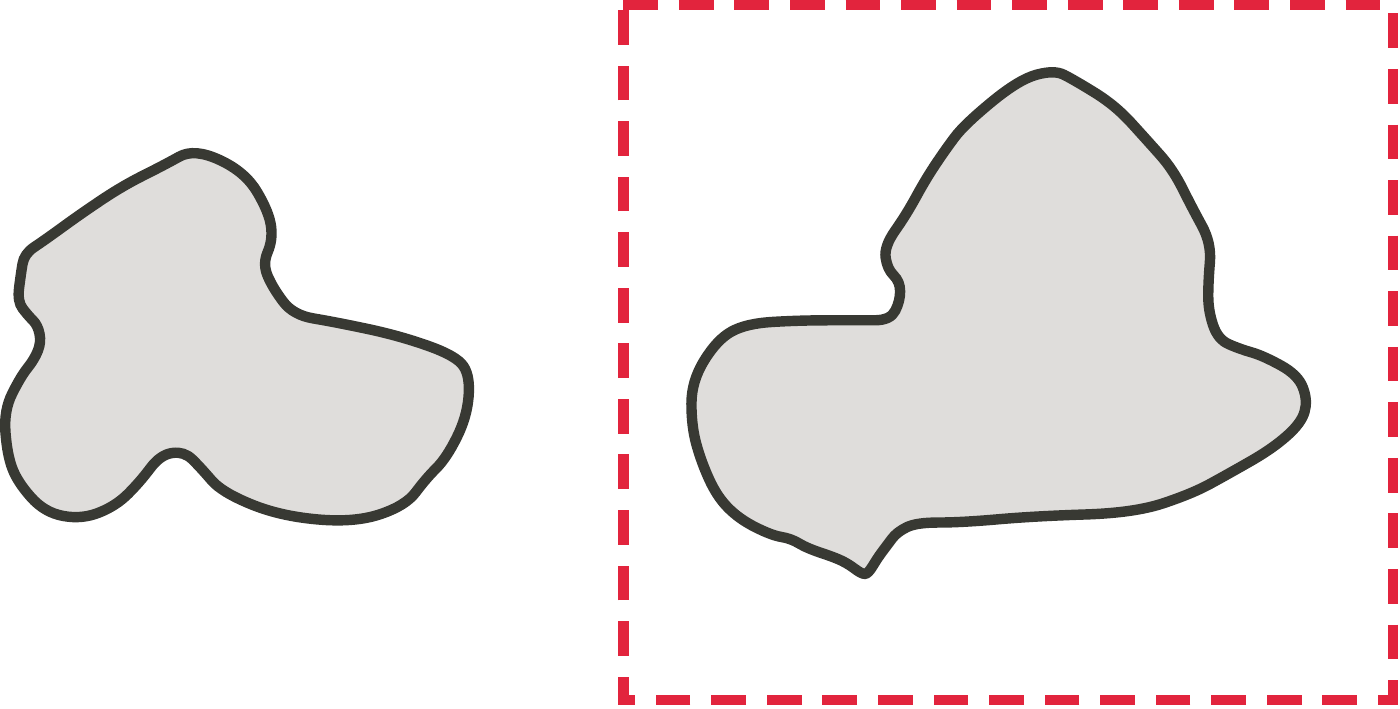}
\par\end{centering}

\caption{\label{fig:stress-schematic}Schematic depiction of two objects whose
Casimir interaction is desired. One computational method involves
integrating a mean stress tensor around some closed surface (dashed
red line) surrounding an object, yielding the force on that object.}

\end{figure}

In practice, one often wants to know the Casimir force (or torque)
on an object rather than the energy density. In this case, instead
of integrating an electromagnetic energy density over the volume,
one can integrate an electromagnetic \emph{stress tensor} over a surface
enclosing the object in question, schematically depicted in Fig.~\ref{fig:stress-schematic}~\cite{Lifshitz80}.%
 The mean stress tensor for the Casimir force is~\cite{Lifshitz80}:
\begin{multline}
\langle T_{jk}(\vec{x})\rangle_{\omega}=\varepsilon(\vec{x},\omega)\left[\langle E_{j}(\vec{x})E_{k}(\vec{x})\rangle_{\omega}-\frac{\delta_{jk}}{2}\sum_{\ell}\langle E_{\ell}(\vec{x})^{2}\rangle_{\omega}\right]\\
+\mu(\vec{x},\omega)\left[\langle H_{j}(\vec{x})H_{k}(\vec{x})\rangle-\frac{\delta_{jk}}{2}\sum_{\ell}\langle H_{\ell}(\vec{x})^{2}\rangle_{\omega}\right].\label{eq:stress}\end{multline}
As above, the field correlation functions are expressed in terms of
the classical Green's function, and the integral of the contributions
over all $\omega$ is Wick-rotated to imaginary frequencies $\omega=i\xi$:
\begin{equation}
\vec{F}=\int_{0}^{\infty}d\xi\oiint_{\mbox{surface}}\langle\mathbf{T}(\vec{x})\rangle_{i\xi}\cdot d\vec{S},\label{eq:stress-integral}\end{equation}
with (zero-temperature) correlation functions \begin{equation}
\langle E_{j}(\vec{x})E_{k}(\vec{x})\rangle_{i\xi}=\frac{\hbar}{\pi}\xi^{2}G_{jk}^{E}(i\xi;\vec{x},\vec{x}),\label{eq:EE-imag}\end{equation}
\begin{equation}
\langle H_{j}(\vec{x})H_{k}(\vec{x})\rangle_{i\xi}=\frac{\hbar}{\pi}\xi^{2}G_{jk}^{H}(i\xi;\vec{x},\vec{x}),\label{eq:HH-imag}\end{equation}
corresponding to the fields on the stress-integration surface in response
to currents placed on that surface. To compute a Casimir torque around
an origin $\vec{r}$, one instead uses $(\vec{x}-\vec{r})\times\langle\mathbf{T}(\vec{x})\rangle_{i\xi}\cdot d\vec{S}$~\cite{Rodriguez08:PRL}.

The derivation of this stress tensor (\ref{eq:stress}) is not as
straightforward as it might at first appear. If the stress-integration
surface lies entirely in vacuum $\varepsilon\approx\varepsilon_{0}$
and $\mu\approx\mu_{0}$, then one can interpret (\ref{eq:stress})
as merely the ordinary EM stress tensor from the microscopic Maxwell
equations~\cite{Jackson98}, albeit integrated over fluctuations.
If the stress-integration surface lies in a dispersive/dissipative
medium such as a fluid, however, then the classical EM stress tensor
is well known to be problematic~\cite{Jackson98} and (\ref{eq:stress})
may seem superficially incorrect. However, it turns out that these
problems disappear in the context of thermodynamic equilibrium, where
a more careful free-energy derivation of the Casimir force from fluctuations
indeed results in equation~(\ref{eq:stress})~\cite{Lifshitz80,Pitaevski06},%
\footnote{If compressibility of the fluid and the density-dependence of $\varepsilon$
is not neglected, then there is an additional $\partial\varepsilon/\partial\rho$
term in (\ref{eq:stress}) resulting from fluctuations in the density~$\rho$~\cite{Lifshitz80}.%
} which has also proved consistent with experiments~\cite{Munday07,Munday09}.
Note also that, while (\ref{eq:stress}) assumes the special case
of isotropic media at $\vec{x}$, it can still be used to evaluate
the force on objects made of anisotropic materials, as long as the
stress-integration surface lies in an isotropic medium (e.g. vacuum
or most fluids).

This formulation is especially important for methods that use an iterative
solver for the Green's functions as discussed below, because it only
requires solving for the response to currents on the stress-integration
surface, rather than currents at every point in space to integrate
the energy density, greatly reducing the number of right-hand sides
to be solved for the linear equation (\ref{eq:Green-imag})~\cite{Rodriguez07:PRA};
additional reductions in the number of right-hand sides are described
in Sec.~\ref{sub:spatial-integration-basis}.

\subsection{Finite-difference frequency-domain (FDFD)}

In order to determine the Casimir energy or force, one evaluates the
Green's function by solving eq.~(\ref{eq:Green-imag}) and then integrates
the appropriate energy/force density over volume/surface and over
the imaginary frequency $\xi$. The central numerical problem is then
the determination of the Green's function by solving a set of linear
equations corresponding to (\ref{eq:Green-imag}), and probably the
simplest approach is based on a finite-difference (FD) basis: space
is divided into a uniform grid with some resolution $\Delta x$, derivatives
are turned into differences, and one solves (\ref{eq:Green-imag})
by some method for every desired right-hand side. In classical EM
(typically finding the field from a given current at real $\omega$),
this is known as a finite-difference frequency-domain (FDFD) method,
and has been widely used for many years~\cite{Christ87,Yasumoto05}.

For example, in one dimension for $z$-directed currents/fields with
$\mu=1$, (\ref{eq:Green-imag}) becomes $\left(-\frac{d^{2}}{dx^{2}}+\xi^{2}\right)\varepsilon G_{zz}^{E}=\delta(x-x')\hat{z}$.
If we approximate $G_{zz}^{E}(n\Delta x,x')\approx G_{n}$, then the
corresponding finite-difference equation, with a standard center-difference
approximation for $d^{2}/dx^{2}$~\cite{strikwerda89}, is \begin{equation}
-\frac{G_{n+1}-2G_{n}+G_{n-1}}{\Delta x^{2}}+\xi^{2}\varepsilon_{n}G_{n}=\frac{\delta_{nn'}}{\Delta x},\label{eq:Green-imag-1d-FD}\end{equation}
replacing the $\delta(x-x')$ with a discrete equivalent at $n'$.
Equation~(\ref{eq:Green-imag-1d-FD}) is a tridiagonal system of
equations for the unknowns $G_{n}$. More generally, of course, one
has derivatives in the $y$ and $z$ directions and three unknown
$\vec{G}$ (or $\vec{E}$) components to determine at each grid point.
As mentioned in Sec.~\ref{sub:Choices-of-basis}, it turns out that
accurate center-difference approximations for the $\nabla\times\nabla\times$
operator in three dimensions are better suited to a {}``staggered''
grid, called a \emph{Yee} grid~\cite{Taflove00,Christ87}, in which
different field components are discretized at points slightly offset
in space: e.g., $E_{x}([n_{x}+\frac{1}{2}]\Delta x,n_{y}\Delta y,n_{z}\Delta z)$,
$E_{y}(n_{x}\Delta x,[n_{y}+\frac{1}{2}]\Delta y,n_{z}\Delta z)$,,
and $E_{z}(n_{x}\Delta x,n_{y}\Delta y,[n_{z}+\frac{1}{2}]\Delta z)$
for the $\vec{E}$ field components. Note that any arbitrary frequency
dependence of $\varepsilon$ is trivial to include, because in frequency
domain one is solving each $\xi$ separately, and a perfect electric
conductor is simply the $\varepsilon(i\xi)\to\infty$ limit.

One must, of course, somehow truncate the computational domain to
a finite region of space in order to obtain a finite number $N$ of
degrees of freedom. There are many reasonable ways to do this because
Casimir interactions are rapidly decaying in space (force $\sim1/a^{d+1}$
or faster with distance $a$ in $d$ dimensions, at least for zero
temperature). One could simply terminate the domain with Dirichlet
or periodic boundary conditions, for example, and if the cell boundaries
are far enough away from the objects of interest then the boundary
effects will be negligible (quite different from classical EM problems
at real $\omega$!)~\cite{Rodriguez07:PRA}. In classical EM, one
commonly uses the more sophisticated approach of a \textbf{perfectly
matched layer} (PML), an artificial reflectionless absorbing material
placed adjacent to the boundaries of the computational domain to eliminate
outgoing waves~\cite{Taflove00}. Mathematically, a PML in a direction
$x$ is equivalent to a complex {}``coordinate stretching'' $\frac{d}{dx}\to\left(1+i\sigma/\omega\right)^{-1}\frac{d}{dx}$
for an artificial PML {}``conductivity'' $\sigma(x)>0$~\cite{Taflove00,Chew94,ZhaoCa96,Teixeira98},
where the $1/\omega$ factor is introduced to give an equal attenuation
rate at all frequencies. However, at imaginary frequencies $\omega=i\xi$,
a PML therefore results simply in a \emph{real} coordinate stretching
$\frac{d}{dx}\to\left(1+\sigma/\xi\right)^{-1}\frac{d}{dx}$: in a
PDE with decaying, non-oscillatory solutions (such as the imaginary-$\omega$
Maxwell equations), it is well known that a reasonable approach to
truncating infinite domains is to perform a (real) coordinate transformation
that compresses space far away where the solution is small~\cite{boyd01:book}.
A convenience of Maxwell's equations is that any coordinate transformation
(real or complex) can be converted merely into a change of $\varepsilon$
and $\mu$~\cite{Ward96}, so any PML can be expressed simply as
a change of materials while keeping the same PDE and discretization~\cite{ZhaoCa96,Teixeira98}.

Such a center-difference scheme is nominally second-order accurate,
with discretization errors that vanish as $O(\Delta x^{2})$~\cite{strikwerda89,Taflove00}.
One can also construct higher-order difference approximations (based
on more grid points per difference). As a practical matter, however,
the accuracy is limited instead by the treatment of material interfaces
where $\varepsilon$ changes discontinuously. If no special allowance
is made for these interfaces, the method still converges, but its
convergence rate is reduced by the discontinuity to $O(\Delta x)$~\cite{Ditkowski01,FarjadpourRo06,KottkeFa08,OskooiKo09}
(unless one has $\vec{E}$ polarization completely parallel to all
interfaces so that there is no field discontinuity). There are various
schemes to restore second-order (or higher) accuracy by employing
specialized FD equations at the interfaces~\cite{Ditkowski01,Zhao10},
but an especially simple scheme involves unmodified FD equations with
modified materials: it turns out that, if the discontinuous $\varepsilon$
is smoothed in a particular way (to avoid introducing first-order
errors by the smoothing itself), then second-order accuracy can be
restored~\cite{FarjadpourRo06,KottkeFa08,OskooiKo09}.%
\footnote{Even if the $\varepsilon$ discontinuities are dealt with in this
way, however, one may still fail to obtain second-order accuracy if
the geometry contains sharp corners, which limit the accuracy to $O(\Delta x^{p})$
for some $1<p<2$~\cite{FarjadpourRo06}. This is an instance of
\emph{Darboux's principle}: the convergence rate of a numerical method
is generally limited by the strongest singularity in the solution
that has not been explicitly compensated for~\cite{boyd01:book}.%
}

Given the FD equations, one must must then choose a solution technique
to solve the resulting linear equations $A\vec{x}=\vec{b}$, where
$\vec{x}$ is the Green's function (or field), $\vec{b}$ is the delta-function
(or current) right-hand side, and $A$ is the discretized $\nabla\times\mu^{-1}\nabla\times{}+\xi^{2}\varepsilon$
operator. Note that $A$ is a real-symmetric positive-definite matrix
at imaginary frequencies, as discussed in Sec.~\ref{sub:imaginary-Green}.
Because $A$ is sparse {[}only $O(N)$ nonzero entries{]}, one can
utilize a sparse-direct Cholesky factorization $A=R^{T}R$ ($R$ is
upper-triangular)~\cite{Davis06} (for which many software packages
are available~\cite{bai00}). Given this factorization, any right-hand
side can be solved quickly by backsubstitution, so one can quickly
sum the energy density over all grid points (essentially computing
the trace of $A^{-1}$) to find the Casimir energy, or alternatively
sum the stress tensor over a stress-integration surface to find the
force. Precisely such a sparse-direct FD method for the Casimir energy
was suggested by~Pasquali and Maggs~\cite{Pasquali09}, albeit derived
by a path-integral $\log\det$ expression that is mathematically equivalent
to summing the energy density (\ref{eq:U-mean-E})~\cite{MiltonWa10:arxiv}.
The alternative is an iterative technique, and in this case $A$'s
Hermitian definiteness means that an ideal Krylov method, the conjugate-gradient
method~\cite{Trefethen97,barrett94} can be employed~\cite{Rodriguez07:PRA}.
The conjugate-gradient method requires $O(N)$ storage and time per
iteration, and in the absence of preconditioning requires a number
of iterations in $d$ dimensions proportional to the diameter $O(N^{1/d})$
of the grid for each right-hand side~\cite{Golub96}. The stress-tensor
approach reduces the number of right-hand sides to be solved compared
to energy-density integration: one only needs to evaluate the Green's
function for sources on a stress-integration surface, which has $O(N^{\frac{d-1}{d}})$
points in $d$ dimensions. This gives a total time complexity of $O(N)\cdot O(N^{1/d})\cdot O(N^{\frac{d-1}{d}})=O(N^{2})$
for an unpreconditioned iterative method; an ideal multigrid preconditioner
can in principle reduce the number of iterations to $O(1)$~\cite{Zhu06,Trottenberg01}
(when $N$ is increased by improving spatial resolution), yielding
an $O(N^{2-\frac{1}{d}})$ time complexity. Substantial further improvements
are obtained by realizing that one does not, in fact, need to sum
over every point on the stress-integration surface, instead switching
to a different spatial integration scheme described in Sec.~\ref{sub:spatial-integration-basis}.

\subsection{Boundary-element methods (BEMs)\label{sub:BEM-stress}}

\begin{figure}[t]

\begin{centering}
\includegraphics[width=0.5\columnwidth]{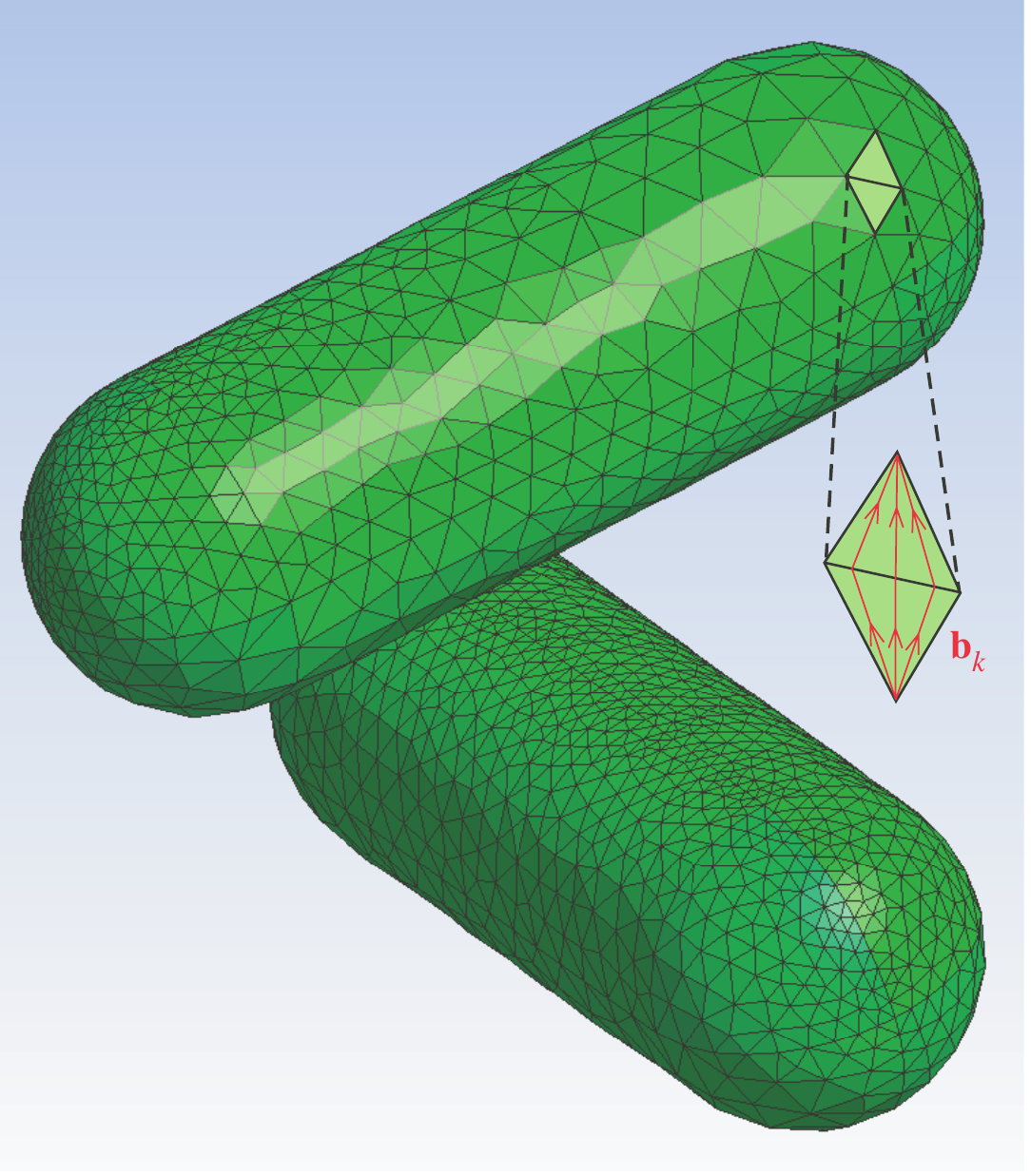}
\par\end{centering}

\caption{\label{fig:crossed-capsules}Example triangular mesh of the surfaces
of two objects for a BEM solver~\cite{ReidRo09}. Associated with each edge $k$ is
an {}``RWG'' basis function $\vec{b}_{k}$~\cite{RWG82}, schematically
represented in the inset, which vanishes outside the adjacent two
triangles.}

\end{figure}

In some sense, a volume discretization such as an FD method is too
general: in most physical situations, the medium is piecewise-constant,
and one might want to take advantage of this fact. In particular,
for the basic problem of finding the field in response to a current
source at a given frequency, one can instead use a surface integral-equation
approach: the unknowns are \emph{surface currents} on the interfaces
between homogeneous materials, and one solves for the surface currents
so that the total field (source + surface currents) satisfies the
appropriate boundary conditions at the interfaces~\cite{chew01,Jin02,Volakis01}.
For example, in the case of a perfect electric conductor, the surface-current
unknowns can be the physical electric currents $\vec{J}$ at the interface,
and the boundary condition is that of vanishing tangential $\vec{E}$
field.%
\footnote{This is known as an \emph{electric-field integral equation} (EFIE);
one can also express the equations for perfect conductors in terms
of boundary conditions enforced on \emph{magnetic} fields (\emph{M}FIE)
or some linear \emph{combination} of the two (\emph{C}FIE), and the
most effective formulation is still a matter of debate~\cite{EpsteinGr09}.%
} In the case of permeable media $\varepsilon$ and $\mu$, the physical
(bound) currents are volumetric within the medium {[}e.g., the electric
bound current is $\vec{J}=-i\omega(\varepsilon-\varepsilon_{0})\vec{E}${]},
not surface currents~\cite{Jackson98}. However, it turns out that
one can introduce \emph{fictitious} surface electric and magnetic
currents at all interfaces to provide enough degrees of freedom to
satisfy the boundary condition of continuous tangential $\vec{E}$
and $\vec{H}$, and thus to fully solve Maxwell's equations. The application
of this \emph{equivalence principle}%
\footnote{The idea of solving scattering problems by introducing fictitous boundary
currents had its origins~\cite{Love01,Schelkunoff36,StrattonChu39,Rengarajan00}
many years before its application to BEM by Harrington~\cite{Harrington89}
and subsequent refinements.%
} to obtain surface integral equations for BEM is known as the PMCHW
approach (Poggio, Miller, Chang, Harrington, and Wu)~\cite{Harrington89,UmashankarTa86,MedgyesiPu94}.
In either case, one has surface (electric and/or magnetic) currents
$\vec{J}_{s}$, plus an external current source $\vec{J}$ {[}e.g.,
the right-hand side of (\ref{eq:Green-real}){]}, so one can express
the $\vec{E}$ or $\vec{H}$ field at any point $\vec{x}$ as a convolution
of $\vec{J}+\vec{J}_{s}$ with the \emph{analytically known} Green's
function $\mathbf{G}_{0}(\vec{x}-\vec{x}')$ of the corresponding
\emph{homogeneous} medium at $\vec{x}$. In BEM, one expresses $\vec{J}_{s}$,
in turn, as a sum of \emph{localized} basis functions $\vec{b}_{k}$
associated with some discrete mesh approximation of the surface. For
example, Fig.~\ref{fig:crossed-capsules} depicts a standard triangular-type
mesh of two objects, where there is a localized basis function $\vec{b}_{k}$
(inset) associated with each \emph{edge} of this mesh such that $\vec{b}_{k}$
is nonzero only on the adjacent two triangles~\cite{RWG82}; this
is the RWG basis mentioned in Sec.~\ref{sub:Choices-of-basis}.%
 Abstractly, the resulting equations for the fields could then be
written in the following form: \begin{equation}
\mbox{field}(\vec{x})=\mathbf{G}_{0}\ast(\vec{J}+\vec{J}_{s})=\mathbf{G}_{0}\ast\vec{J}+\sum_{k=1}^{N}\mathbf{G}_{0}\ast\vec{b}_{k}c_{k},\label{eq:BEM-abstract}\end{equation}
where $\mathbf{G}_{0}\ast{}$ denotes convolution with the (dyadic)
analytical Green's function of the homogeneous medium at $\vec{x}$,
and $c_{k}$ are the unknown coefficients of each basis function.%
\footnote{Technically, only currents from surfaces bordering the medium of $\vec{x}$
contribute to this sum.%
} (More generally, $\mathbf{G}_{0}\ast\vec{J}$ could be replaced by
any arbitrary incident field, regardless of how it is created.) In
a Galerkin method (see Sec.~\ref{sub:Choices-of-basis}), one obtains
$N$ equations for the $N$ unknowns $c_{k}$ by taking the inner
product of both sides of this equation (substituted into the appropriate
boundary condition) with the same basis functions $\vec{b}_{j}$ (since
they work just as well as a basis for the tangential field as for
the tangential surface currents). This ultimately results in a set
of linear equations $A\vec{c}=\vec{d}$, where the matrix $A$ multiplying
the unknown coefficients $c_{k}$ is given by \begin{equation}
A_{jk}=\iint\bar{\vec{b}}_{j}(\vec{x})\cdot\mathbf{G}_{0}(\vec{x}-\vec{x}')\cdot\vec{b}_{k}(\vec{x}')\, d^{2}\vec{x}\, d^{2}\vec{x}'.\label{eq:BEM-matrix}\end{equation}
{[}For the case of a perfect conductor with a vanishing tangential~$\vec{E}$,
the right-hand-side~$\vec{d}$ is given by $d_{j}=-\langle\vec{b}_{j},\mathbf{G}_{0}\ast\vec{J}\rangle=-\iint\bar{\vec{b}}_{j}(\vec{x})\cdot\mathbf{G}_{0}(\vec{x}-\vec{x}')\cdot\vec{J}(\vec{x}')\, d^{2}\vec{x}\, d^{2}\vec{x}'$.{]}
One then solves the linear system for the unknown coefficients $c_{k}$,
and hence for the unknown surface currents $\vec{J}_{s}$. Implementing
this technique is nontrivial because the $A_{jk}$ integrands (\ref{eq:BEM-matrix})
are singular for $j=k$ or for $j$ adjacent to $k$, necessitating
specialized quadrature techniques for a given form of $\mathbf{G}_{0}$~\cite{Taylor03,TongCh06},
but substantial guidance from the past decades of literature on the
subject is available.

Given these currents, one can then evaluate the electric or magnetic
field at any point $\vec{x}$, not just on the surface, by evaluating
eq.~(\ref{eq:BEM-abstract}) at that point. In particular, one can
evaluate the field correlation functions via the fluctuation-dissipation
theorem (\ref{eq:Ecorr-real}): $\langle E_{j}(\vec{x})E_{k}(\vec{x})\rangle_{\omega}$
is given in terms of the electric field in the $j$ direction at $\vec{x}$
from a delta-function current $\vec{J}$ in the $k$ direction at
$\vec{x}$. Of course, as noted previously, this is infinite because
the $\vec{G}_{0}\ast\vec{J}$ term (the field from the delta function)
blows up at $\vec{x}$, but in the Casimir case one is only interested
in the \emph{change} of the correlation functions due to the geometry---so,
one can use the standard trick~\cite{Lifshitz80} of subtracting
the vacuum contribution $\mathbf{G}_{0}\ast\vec{J}$ and only computing
the surface-current contribution $\mathbf{G}_{0}\ast\vec{J}_{s}$
to the field at $\vec{x}$. In this way, one can compute the stress
tensor, the energy density, and so on, as desired.

As explained in Sec.~\ref{sec:complex-frequency}, the integral of
contributions over all frequencies is best performed at imaginary
frequencies, so all of the above must use $\omega=i\xi$. This only
has the effect of Wick-rotating the homogeneous-medium dyadic Green's
function $\mathbf{G}_{0}$ to the $\sim e^{-\xi|\vec{x}-\vec{x}'|}/|\vec{x}-\vec{x}'|$
imaginary-frequency Green's function. This makes the problem \emph{easier},
in principle. First, the exponential decay cuts off long-range interactions,
making fast-solver techniques (see Sec.~\ref{sub:Solution-techniques})
potentially even more effective. Second, the matrix $A$ is now real-symmetric
and positive-definite, which allows the use of more efficient linear
solvers as noted previously. Fortunately, the $1/|\vec{x}-\vec{x}'|$
singularity of $\mathbf{G}_{0}$ is the same at real and imaginary
frequencies, allowing existing techniques for the integration of (\ref{eq:BEM-matrix})
to be leveraged.

At first glance, this approach seems most straightforwardly applicable
to the stress-tensor technique, as suggested in Ref.~\cite{Rodriguez07:PRA}:
one uses the BEM solution to evaluate the mean stress tensor $\langle\mathbf{T}\rangle$
on any integration surface around a body, integrating via some quadrature
technique to obtain the force. If one uses a dense-direct solver (when
$N$ is not too big), the Cholesky factorization of $A$ can be computed
once for a given $\xi$ and then many right-hand sides can be solved
quickly via backsubstitution~\cite{Trefethen97} in order to integrate
$\langle\mathbf{T}\rangle$ over the stress-integration surface. Precisely
such a dense-direct BEM stress-tensor method was recently demonstrated
to compute Casimir forces in two and three dimensions~\cite{XiongCh09,XiongTo10:arxiv}.
As described in Sec.~\ref{sub:Solution-techniques}, fast-solver
techniques can be applied to multiply $A$ by a vector in $O(N\log N)$
time with $O(N)$ storage; given a good preconditioner, this implies
that an iterative method such as conjugate-gradient (applicable since
$A$ is real-symmetric positive-definite) could find $\langle\mathbf{T}\rangle$
at a single $\vec{x}$ and $\xi$ in $O(N\log N)$ time. The remaining
question is the number of points required for the surface integral
of $\langle\mathbf{T}\rangle$, which depends on \emph{why} one is
increasing $N$: either to increase accuracy for a fixed geometry
or to increase the complexity of the geometry for a fixed accuracy.
In the former case, the smoothness of $\langle\mathbf{T}\rangle$
in $\vec{x}$ means that exponentially convergent quadrature techniques
are applicable, which converge much faster than the (polynomial) BEM
basis for the surface currents, so that ultimately the number of stress-quadrature
points%
\footnote{Numeric integration (\emph{quadrature}) approximates an integral $\int f(x)dx$
by a sum $\sum_{i}f(x_{i})w_{i}$ over quadrature points $x_{i}$
with weights $w_{i}$. There are many techniques for the selection
of these points and weights, and in general one can obtain an error
that decreases exponentially fast with the number of points for analytic
integrands~\cite{StroudSe66,piessens83,boyd01:book,Trefethen08}.
Multidimensional quadrature, sometimes called \emph{cubature}, should
be used to integrate the stress tensor over a 2d surface, and numerous
schemes have been developed for low-dimensional cubature~\cite{Cools02-review,Cools03}
(including methods that adaptively place more quadrature points where
they are most needed~\cite{Berntsen91}). For spherical integration
surfaces (or surfaces that can be smoothly mapped to spheres), specialized
methods are available~\cite{Atkinson05,LeGia08}.%
} should be independent of $N$ and the overall complexity becomes
$O(N\log N)$. In the latter case, for a fixed accuracy and increasingly
complex geometry (or smaller feature sizes), it appears likely that
the number of stress-quadrature points will increase with $N$, but
detailed studies of this scaling are not yet available.

It turns out that this BEM approach is closely related to the BEM
path-integral approach described in Sec.~\ref{sub:BEM-path}. Both
approaches end up solving linear equations with exactly the same matrix
$A$ of eq.~(\ref{eq:BEM-matrix}), with the same degrees of freedom.
The path-integral approach shows, however, that this same matrix can
be applied to compute the Casimir interaction energy as well as the
force, with comparable computational cost for dense solvers. Moreover,
as explained below, expressing the force in terms of the derivative
of the path-integral energy results in a trace expression that is
conceptually equivalent to integrating a stress tensor over the surface
of an object, where the number of {}``quadrature points'' is now
exactly equal to $N$. An unanswered question, at this point, is whether
a fast solver can be more efficiently (or more easily) exploited in
the stress-tensor approach or in the path-integral approach.

\subsection{Other possibilities: FEM and spectral methods}

There are of course, many other frequency-domain techniques from classical
EM that could potentially be used to solve for the Green's function
and hence the energy/force density. For example, one could use spectral
integral-equation methods, such as multipole expansions for spheres
and cylinders~\cite{Yasumoto05}, to compute responses to currents,
although the advantages of this approach compared to the spectral
path-integral approach in Sec.~\ref{sub:Spectral-path-integral}
are unclear. One can also solve the PDE formulation of the Green's
function (\ref{eq:Green-imag}) using a finite-element (FEM) approach
with some general mesh; in principle, existing FEM techniques from
classical EM~\cite{Volakis01,chew01,Jin02,Zhu06} are straightforwardly
applicable. One subtlety that arises in FEM methods with a nonuniform
resolution is the regularization, however~\cite{Rodriguez07:PRA}.
In principle, as mentioned above, one needs to subtract the vacuum
Green's function contribution from the field correlation functions
in order to get a physical result {[}since the vacuum Green's function
$\mathbf{G}(\vec{x},\vec{x}')$ diverges as $\vec{x}'\to\vec{x}$,
although the divergence is cut off by the the finite mesh resolution{]}.
With a uniform mesh, this vacuum contribution is the same everywhere
in the mesh and hence automatically integrates to zero in the force
(when the stress tensor is integrated over a closed surface or the
energy is differentiated). For a nonuniform mesh, however, the vacuum
contribution varies at different points in space with different resolution,
so some {}``manual'' regularization seems to be required (e.g.,
subtracting a calculation with the same mesh but removing the objects).
These possibilities currently remain to be explored for Casimir physics.

\subsection{Finite-difference time-domain (FDTD) methods\label{sub:FDTD}}

Casimir effects are fundamentally broad-bandwidth, integrating contributions
of fluctuations at all frequencies (real or imaginary), although the
imaginary-frequency response is dominated by a limited range of imaginary
frequencies. In classical EM, when a broad-bandwidth response is desired,
such as a transmission or reflection spectrum from some structure,
there is a well-known alternative to computing the contributions at
each frequency separately---instead, one can simulate the same problem
in \emph{time}, Fourier-transforming the response to a short pulse
excitation in order to obtain the broad-bandwidth response in a single
\emph{time-domain }simulation~\cite{JoannopoulosJo08-book,Oskooi10:Meep}.
The same ideas are applicable to the Casimir problem, with a few twists,
yielding a practical method~\cite{RodriguezMc09:PRA,McCauleyRo10:PRA}
that allows Casimir calculations to exploit off-the-shelf time-domain
solvers implementing the standard \emph{finite-difference time-domain}
(FDTD) method~\cite{Taflove00}. There are two key components of
this approach~\cite{RodriguezMc09:PRA}: first, converting the frequency
integral to a time integral and, second, finding a time-domain equivalent
of the complex-frequency idea from Sec.~\ref{sec:complex-frequency}.

As reviewed above, the mean fluctuations in the fields, such as $\langle E^{2}(\vec{x})\rangle_{\omega}$,
can be expressed in terms of the fields at $\vec{x}$ from a frequency-$\omega$
current at $\vec{x}$. If, instead of a frequency-$\omega$ current,
one uses a current with $\delta(t)$ time dependence, it follows by
linearity of (\ref{eq:Green-real}) that the Fourier transform of
the resulting fields must yield exactly the same $\langle E^{2}(\vec{x})\rangle_{\omega}$.
Roughly, the procedure could be expressed as follows: First, we compute
some function $\Gamma(t)$ of the time-domain fields from a sequence
of simulations with $\delta(t)$ sources, e.g. where $\Gamma(t)$
is the result of spatially integrating the fields making up the mean
stress tensor $\langle\mathbf{T}(\vec{x})\rangle$ {[}noting that
each point $\vec{x}$ involves several separate $\delta(t)$-response
simulations{]}. Second, we Fourier transform $\Gamma(t)$ to obtain
$\tilde{\Gamma}(\omega)$. Third, we obtain the force (or energy,
etcetera) by integrating $\int\tilde{\Gamma}(\omega)\tilde{g}(\omega)d\omega$
with appropriate frequency-weighting factor $\tilde{g}(\omega)$ (which
may come from the frequency dependence of $\varepsilon$ in $\langle\mathbf{T}\rangle$,
a Jacobian factor from below, etcetera). At this point, however, it
is clear that the Fourier transform of $\Gamma$ was entirely unnecessary:
because of the unitarity of the Fourier transform (the Plancherel
theorem), $\int\tilde{\Gamma}(\omega)\tilde{g}(\omega)d\omega=\int\Gamma(t)g(-t)dt$.
That is, we can compute the force (or energy. etcetera) by starting
with $\delta(t)$ sources and simply integrating the response $\Gamma(t)$
in time (accumulated as the simulation progresses) multiplied by some
(precomputed, geometry-independent) kernel $g(t)$ (which depends
on temperature if the $\coth$ factor is included for $T>0$). The
details of this process, for the case of the stress tensor, are described
in Refs.~\cite{RodriguezMc09:PRA,McCauleyRo10:PRA}.

Although it turns out to be possible to carry out this time-integration
process as-is, we again find that a transformation into the complex-frequency
plane is desirable for practical computation (here, to reduce the
required simulation time)~\cite{RodriguezMc09:PRA}. Transforming
the \emph{frequency} in a \emph{time}-domain method, however, requires
an indirect approach. The central observation is that, in the equation
(\ref{eq:Green-real}) for the electric-field Green's function $\mathbf{G}^{E}$
, the frequency only appears explicitly in the $\omega^{2}\varepsilon$
term, together with $\varepsilon$. So, any transformation of $\omega$
can equivalently be viewed as a transformation of $\varepsilon$.
In particular suppose that we wish to make some transformation $\omega\to\omega(\xi)$
to obtain an $\omega$ in the upper-half complex plane, where $\xi$
is a real parameter (e.g. $\omega=i\xi$ for a Wick rotation). Equivalently,
we can view this as a calculation at a \emph{real} frequency $\xi$
for a transformed \emph{complex material}: $\omega^{2}\varepsilon(\omega,\vec{x})\to\xi^{2}\varepsilon_{c}(\xi,\vec{x})$
where the transformed material is~\cite{RodriguezMc09:PRA,RodriguezMc09:analog}
\begin{equation}
\varepsilon_{c}(\xi,\vec{x})=\frac{\omega^{2}(\xi)}{\xi^{2}}\varepsilon(\omega(\xi),\vec{x}).\label{eq:epsilon-contour}\end{equation}
For example, a Wick rotation $\omega\to i\xi$ is equivalent to operating
at a real frequency $\xi$ with a material $\varepsilon(\omega)\to-\varepsilon(i\xi)$.
However, at this point we run into a problem: multiplying $\varepsilon$
by $-1$ yields exponentially growing solutions at negative frequencies~\cite{RodriguezMc09:PRA,RodriguezMc09:analog},
and this will inevitably lead to exponential blowup in a time-domain
simulation (which cannot avoid exciting negative frequencies, if only
from roundoff noise). In order to obtain a useful time-domain simulation,
we must choose a contour $\omega(\xi)$ that yields a \emph{causal},
\emph{dissipative} material $\varepsilon_{c}$, and one such choice
is $\omega(\xi)=\xi\sqrt{1+i\sigma/\xi}$ for any constant $\sigma>0$~\cite{RodriguezMc09:PRA,RodriguezMc09:analog}.
This yields $\varepsilon_{c}=(1+i\sigma/\omega)\varepsilon$, where
the $i\sigma/\omega$ term behaves exactly like an artificial \emph{conductivity}
added everywhere in space. In the frequency-domain picture, we would
say from Sec.~\ref{sec:complex-frequency} that this $\omega(\xi)$
contour will improve the computation by moving away from the real-$\omega$
axis, transforming the frequency integrand into something exponentially
decaying and less oscillatory. In the time-domain picture, the $\sigma$
term adds a \emph{dissipation} everywhere in space that causes $\Gamma(t)$
to \emph{decay exponentially in time}, allowing us to truncate the
simulation after a short time. As long as we include the appropriate
Jacobian factor $\frac{d\omega}{d\xi}$ in our frequency integral,
absorbing it into $g(t)$, we will obtain the \emph{same result} in
a much shorter time. The computational details of this transformation
are described in Refs.~\cite{RodriguezMc09:PRA,McCauleyRo10:PRA}.
More generally, this equivalence between the Casimir force and a relatively
narrow-bandwidth real-frequency response of a dissipative system potentially
opens other avenues for the understanding of Casimir physics~\cite{RodriguezMc09:analog}.

The end result is a computational method for the Casimir force in
which one takes an off-the-shelf time-domain solver (real time/frequency),
adds an artificial conductivity $\sigma$ everywhere, and then accumulates
the response $\Gamma(t)$ to short pulses multiplied by a precomputed
(geometry independent) kernel $g(t)$. The most common time-domain
simulation technique in classical EM is the FDTD method~\cite{Taflove00}.
Essentially, FDTD works by taking the same spatial Yee discretization
as in the FDFD method above, and then also discretizing time with
some time step $\Delta t$. The fields are then marched through time
in steps of $\Delta t$, where each time step requires $O(N)$ work
for $N$ spatial grid points. Because the complex-$\omega$ contour
is implemented entirely as a choice of materials $\varepsilon_{c}$,
existing FDTD software can be used without modification to compute
Casimir forces, and one can exploit powerful existing software implementing
parallel calculations, various dimensionalities and symmetries, general
dispersive and anisotropic materials, PML absorbing boundaries, and
techniques for accurate handling of discontinuous materials. One such
FDTD package is available as free/open-source software from our group~\cite{Oskooi10:Meep},
and we have included built-in facilities to compute Casimir forces~\cite{casimir-wiki}.

\subsection{Accelerating FD convergence\label{sub:spatial-integration-basis}}

Finally, we should mention a few techniques that accelerate the convergence
and reduce the computational cost of the finite-difference approaches.
These techniques are not \emph{necessary} for convergence, but they
are simple to implement and provide significant efficiency benefits.

The simplest technique is extrapolation in $\Delta x$: since the
convergence rate of the error with the spatial resolution $\Delta x$
is generally known \emph{a priori}, one can fit the results computed
at two or more resolutions in order to extrapolate to $\Delta x\to0$.
The generalization of this approach is known as \emph{Richardson extrapolation}~\cite{Press92},
and it can essential increases the convergence order cheaply, e.g.,
improving $O(\Delta x)$ to $O(\Delta x^{2})$~\cite{Werner07}.

Second, suppose one is computing the force between two objects $A$
and $B$ surrounded by a homogeneous medium. If one of the objects,
say $B$, is removed, then (in principle) there should be no net remaining
force on $A$. However, because of discretization asymmetry, a computation
with $A$ alone will sometimes still give a small net force, which
converges to zero as $\Delta x\to0$. If this {}``error'' force
is subtracted from the $A$--$B$ force calculation, it turns out
that the net error is reduced. More generally, the error is greatly
reduced if one computes the $A$--$B$ force and then subtracts the
{}``error'' forces for $A$ alone and for $B$ alone, tripling the
number of computations but greatly reducing the resolution that is
required for an accurate result~\cite{Rodriguez07:PRA}.

Third, when integrating the stress tensor $\langle\mathbf{T}(\vec{x})\rangle_{i\xi}$
over $\vec{x}$ to obtain the net force (\ref{eq:stress-integral}),
the most straightforward technique in FD is to simply sum over all
the grid points on the integration surface---recall that each point
$\vec{x}$ requires a linear solve (a different right-hand side) in
frequency domain, or alternatively a separate time-domain simulation
(a separate current pulse). This is wasteful, however, because $\langle\mathbf{T}(\vec{x})\rangle_{i\xi}$
is conceptually smoothly varying in space---if one could evaluate
it at arbitrary points $\vec{x}$ (as is possible in the BEM approach),
an exponentially convergent quadrature scheme could be exploited to
obtain an accurate integral with just a few $\vec{x}$'s. This is
not directly possible in an FD method, but one can employ a related
approach. If the integration surface is a box aligned with the grid,
one can expand the fields on each side of the box in a cosine series
(a discrete cosine transform, or DCT, since space is discrete)---this
generally converges rapidly, so only a small number terms from each
side are required for an accurate integration. But instead of putting
in point sources, obtaining the responses, and expanding the response
in a cosine series, it is equivalent (by linearity) to put in cosine
sources directly instead of point sources. {[}Mathematically, we are
exploiting the fact that a delta function can be expanded in any orthonormal
basis $b_{n}(\vec{x})$ over the surface, such as a cosine series,
via: $\delta(\vec{x}-\vec{x}')=\sum_{n}\bar{b}_{n}(\vec{x}')b_{n}(\vec{x})$.
Substituting this into the right-hand side of (\ref{eq:Green-imag}),
each $b_{n}(\vec{x})$ acts like a current source and $\bar{b}_{n}(\vec{x}')$
scales the result, which is eventually integrated over $\vec{x}'$.{]}
The details of this process and its convergence rate are described
in Ref.~\cite{McCauleyRo10:PRA}, but the consequence is that many
fewer linear systems (fewer right-hand sides) need be solved (either
in frequency or time domain) than if one solved for the stress tensor
at each point individually.

\section{Path integrals and scattering matrices\label{sec:path-integrals}\label{sub:scattering-matrix}}

Another formulation of Casimir interactions is to use a derivation
based on path integrals. Although the path-integral derivation itself
is a bit unusual from the perspective of classical EM, and there are
several slightly different variations on this idea in the literature,
the end result is straightforward: Casimir energies and forces are
expressed in terms of log determinants and traces of classical scattering
matrices~\cite{emig01,maianeto05,emig06,Lambrecht06,Emig07,Rahi07,Kenneth08,MaiaNeto08,Reynaud08,Rahi09:PRD,Lambrecht09},
or similarly the interaction matrices (\ref{eq:BEM-matrix}) that
arise in BEM formulations~\cite{ReidRo09}. Here, we omit the details
of the derivations and focus mainly on the common case of piecewise-homogeneous
materials, emphasizing the relationship of the resulting method to
surface-integral equations from classical EM via the approach in Ref.~\cite{ReidRo09}.

Path integrals relate the Casimir interaction energy $U$ of a given
configuration to a functional integral over all possible vector-potential
fields $\vec{A}$. Assuming piecewise-homogeneous materials, the constraint
that the fields in this path integral must satisfy the appropriate
boundary conditions can be expressed in terms of auxiliary fields
$\vec{J}$ at the interfaces (a sort of Lagrange multiplier)~\cite{LiKardar91}.%
\footnote{Alternatively, the path integral can be performed directly in $\vec{A}$,
resulting in an expression equivalent to the sum over energy density
in Sec.~\ref{sec:fluctuation-dissipation}~\cite{MiltonWa10:arxiv}
and which in an FD discretization reduces in the same way to repeated
solution of the Green's-function diagonal at every point in space~\cite{Pasquali09}.%
} At this point, the original functional integral over $\vec{A}$ can
be performed analytically, resulting in an energy expression involving
a functional integral $Z(\xi)$ over only the auxiliary fields $\vec{J}$
at each imaginary frequency $\xi$~, of the form (at zero temperature):
\begin{equation}
U=-\frac{\hbar c}{2\pi}\int_{0}^{\infty}\log\det\frac{Z(\xi)}{Z_{\infty}(\xi)}d\xi,\label{eq:path-integral-energy}\end{equation}
\begin{equation}
Z(\xi)=\int\mathcal{D}\vec{J}e^{-\frac{1}{2}\iint d^{2}\vec{x}\iint d^{2}\vec{x}'\vec{J}(\vec{x})\cdot\mathbf{G}_{\xi}(\vec{x}-\vec{x}')\cdot\vec{J}(\vec{x}')}.\label{eq:path-integral}\end{equation}
Here, $Z_{\infty}$ denotes $Z$ when the objects are at infinite
separation (non-interacting), regularizing $U$ to just the (finite)
interaction energy (See also the chapter by S.~J.~Rahi et~al.\ in this volume for additional discussion of path integrals and Casimir interactions.) In the case of perfect electric conductors in
vacuum, $\vec{J}$ can be interpreted as a surface current on each
conductor (enforcing the vanishing tangential $\vec{E}$ field), and
$\mathbf{G}_{\xi}$ is the vacuum Green's function in the medium outside
the conductors~\cite{ReidRo09}. For permeable media (finite $\varepsilon$
and $\mu$), it turns out that a formulation closely related to the
standard PMCHW integral-equation model (see Sec.~\ref{sub:BEM-stress})
can be obtained: $\vec{J}$ represents fictitious surface electric
and magnetic currents on each interface (derived from the continuity
of the tangential $\vec{E}$ and $\vec{H}$ fields), with $\mathbf{G}_{\xi}$
again being a homogeneous Green's function (with one $Z$ factor for
each contiguous homogeneous region)~\cite{Reid09-PMCHW}. Alternatively,
because there is a direct correspondence between surface currents
and the outgoing/scattered fields from a given interface, {}``currents''
$\vec{J}$ can be replaced by scattered fields, again related at different
points $\vec{x}$ and $\vec{x}'$ by the Green's function of the homogeneous
medium; this is typically derived directly from a T-matrix formalism~\cite{Emig07,Kenneth08,MaiaNeto08,Rahi09:PRD}.
Here, we will focus on the surface-current viewpoint, which is more
common in the classical-EM integral-equation community.

The path integral~(\ref{eq:path-integral}) is somewhat exotic in
classical EM, but it quickly reduces to a manageable expression once
an approximate (finite) basis $\vec{b}_{k}$ is chosen for the currents
$\vec{J}$. Expanding in this basis, $\vec{J}\approx\sum c_{k}\vec{b}_{k}(\vec{x})$
and the functional integral $\mathcal{D}\vec{J}$ is replaced by an
ordinary integral over the basis coefficients $dc_{1}\cdots dc_{N}$.
Equation~(\ref{eq:path-integral}) is then a Gaussian integral that
can be performed analytically to obtain $Z(\xi)=\#/\sqrt{\det A(\xi)}$
for a proportionality constant $\#$~\cite{ReidRo09}, where $A_{jk}=\int\bar{\vec{b}}_{j}\cdot\vec{G}_{\xi}\cdot\vec{b}_{k}$
is essentially the same as the BEM matrix~(\ref{eq:BEM-matrix}),
albeit here in an arbitrary basis. In the $\log\det$ of~(\ref{eq:path-integral-energy}),
proportionality constants and exponents cancel, leaving: \begin{equation}
U=+\frac{\hbar c}{2\pi}\int_{0}^{\infty}\log\det\left[A_{\infty}(\xi)^{-1}A(\xi)\right]d\xi.\label{eq:path-integral-energy-matrix}\end{equation}
Just as in Sec.~\ref{eq:BEM-matrix}, the use of a real-symmetric
positive-definite homogeneous Green's function $\mathbf{G}_{\xi}$
at imaginary frequencies means that $A(\xi)$ is also real-symmetric
and positive-definite, ensuring positive real eigenvalues and hence
a real $\log\det$. Several further simplifications are possible,
even before choosing a particular basis. For example, let $\vec{p}$
be the position of some object for which the force $\vec{F}$ is desired.
The components $F_{i}$ of the force (in direction $p_{i}$) can then
be expressed directly as a trace~\cite{emig03_2,ReidRo09}: \begin{equation}
F_{i}=-\frac{dU}{dp_{i}}=-\frac{\hbar c}{2\pi}\int_{0}^{\infty}\tr\left(A^{-1}\frac{\partial A}{\partial p_{i}}\right)d\xi.\label{eq:path-integral-force-matrix}\end{equation}
Equivalently, this trace is the sum of eigenvalues $\lambda$ of the
generalized eigenproblem $\frac{\partial A}{\partial p_{i}}\vec{v}=\lambda A\vec{v}$;
again, these $\lambda$ are real because $A$ is real-symmetric positive-definite
and $\partial A/\partial p_{i}$ is real-symmetric. (If dense-direct
solvers are used, computing $A^{-1}\frac{\partial A}{\partial p_{i}}$
via Cholesky factorization is much more efficient than computing eigenvalues,
however~\cite{Trefethen97}.) The matrix $A$ can be further block-decomposed
in the usual case where one is computing the interactions among two
or more disjoint objects (with disjoint surface currents $\vec{J}$).
For example, suppose that one has two objects $1$ and $2$, in which
case one can write \begin{equation}
A=\left(\begin{array}{cc}
A_{11} & A_{12}\\
A_{12}^{\mathrm{T}} & A_{22}\end{array}\right),\label{eq:BEM-matrix-block}\end{equation}
where $A_{11}$ and $A_{22}$ couple currents on each object to other
currents on the same object, and $A_{12}$ and $A_{12}^{\mathrm{T}}=A_{21}$
couple currents on object~1 to object~2 and vice versa. In the limit
of infinite separation for $A_{\infty}$, one obtains $A_{12}\to0$
while $A_{11}$ and $A_{22}$ are unchanged, and one can simplify
the $\log\det$ in (\ref{eq:path-integral-energy-matrix}) to \begin{equation}
\log\det\left[A_{\infty}(\xi)^{-1}A(\xi)\right]=\log\det\left[I-A_{22}^{-1}A_{12}^{\mathrm{T}}A_{11}^{-1}A_{21}\right].\label{eq:logdet-block}\end{equation}
Computationally, only $A_{12}$ depends on the relative positions
of the objects, and this simplification immediately allows several
computations to be re-used if the energy or force is computed for
multiple relative positions.

\subsection{Monte-Carlo path integration}

Before we continue, it should be noted that there also exists a fundamentally
different approach for evaluating a path-integral Casimir formulation.
Instead of reducing the problem to surface/scattering unknowns and
analytically integrating $Z$ to obtain a matrix $\log\det$ expression,
it is possible to retain the original path-integral expression, in
terms of a functional integral over vector potentials $\vec{A}$ in
the volume, and perform this functional integral numerically via Monte-Carlo
methods~\cite{gies03,Gies06:worldline}. This reduces to a Monte-Carlo
integration of an action over all possible closed-loop paths ({}``worldlines''),
discretized into some number of points per path. Because this technique
is so different from typical classical EM computations, it is difficult
to directly compare with the other approaches in this review. Evaluating
its computational requirements involves a statistical analysis of
the scaling of the necessary number of paths and number of points
per path with the desired accuracy and the complexity of the geometry~\cite{Rodriguez07:PRA},
which is not currently available. A difficulty with this technique
is that it has currently only been formulated for scalar fields with
Dirichlet boundary conditions, not for the true Casimir force of vector
electromagnetism.

\subsection{Spectral methods\label{sub:Spectral-path-integral}}

One choice of basis functions $\vec{b}_{k}$ for the path-integral
expressions above is a spectral basis, and (mirroring the history
of integral equations in classical EM) this was the first approach
applied in the Casimir problem. With cylindrical objects, for example,
the natural spectral basis is a Fourier-series $e^{im\phi}$ in the
angular direction $\phi$. For planar surfaces the natural choice
is a Fourier transform, for spheres it is spherical harmonics $Y_{\ell m}$
(or their vector equivalents~\cite{Jackson98}), and for spheroids
there are spheroidal harmonics~\cite{EmigGr09}. Equivalently, instead
of thinking of surface currents expanded in a Fourier-like basis,
one can think of the scattered fields from each object expanded in
the corresponding Fourier-like basis (e.g. plane, cylindrical, or
spherical waves), in which case $A$ relates the incoming to outgoing/scattered
waves for each object; this has been called a {}``scattering-matrix''
or {}``T-matrix'' method and is the source of many pioneering results
for Casimir interactions of non-planar geometries~\cite{Emig07,Kenneth08,MaiaNeto08,Rahi09:PRD}.
Even for nonspherical/spheroidal objects, one can expand the scattered
waves in vector spherical harmonics~\cite{Rahi09:PRD}, and a variety
of numerical techniques have been developed to relate a spherical-harmonic
basis to the boundary conditions on nonspherical surfaces~\cite{KuoTi08}.
These spectral scattering methods have their roots in many classical
techniques for EM scattering problems~\cite{Stratton41,Waterman07} (See also the chapters by S.~J.~Rahi et~al.\ and A.~Lambrecht et~al.\ in this volume for additional discussions of scattering techniques and Casimir interactions.) Here, we
will use the surface-current viewpoint rather than the equivalent
scattered-wave viewpoint.

Many simplifications occur in the interaction matrix $A$ of (\ref{eq:BEM-matrix-block})
for geometries with highly symmetrical objects and a corresponding
spectral basis~\cite{Rahi09:PRD}. Consider, for example, the case
of spherical objects, with surface currents expressed in a vector
spherical-harmonic basis (spherical harmonics for two polarizations~\cite{Rahi09:PRD}).
In the interaction matrix $A_{jk}=\iint\vec{b}_{j}(\vec{x})\cdot\mathbf{G}(\vec{x}-\vec{x}')\cdot\vec{b}_{k}(\vec{x}')$,
the convolution $\int G(\vec{x}-\vec{x}')\cdot\vec{b}_{k}(\vec{x}')$
of a Green's function $\vec{G}$ with $\vec{b}_{k}$ is known analytically:
it is just the outgoing spherical wave produced by a spherical-harmonic
current. If $\vec{b}_{j}$ is another spherical-harmonic current on
the same sphere, then the orthogonality of the spherical harmonics
means that the $\vec{x}$ integral of $\vec{b}_{j}(\vec{x})$ against
the spherical wave is zero unless $j=k$. Thus, the self-interaction
blocks $A_{11}$ and $A_{22}$ of (\ref{eq:BEM-matrix-block}), with
an appropriate normalization, are simply identity matrices. The $A_{12}$
entries are given by the coupling of a spherical wave from $\vec{b}_{k}$
on sphere~2 with a spherical-harmonic basis function $\vec{b}_{j}$
on sphere~1, but again this integral can be expressed analytically,
albeit as an infinite series: the spherical wave from sphere~2 can
be re-expressed in the basis of spherical waves centered on sphere~1
via known translation identities of spherical waves, and as a result
$A_{12}$ takes the form of a {}``translation matrix''~\cite{Rahi09:PRD}.
Furthermore, if there are only two spheres in the problem, then their
spherical harmonics can be expressed with respect to a common $z$
axis passing through the centers of the spheres, and a $Y_{\ell m}$
on sphere~1 will only couple with a $Y_{\ell'm'}$ on sphere~2 if
$m=m'$, greatly reducing the number of nonzero matrix elements. Related
identities are available for coupling cylindrical waves around different
origins, expanding spherical/cylindrical waves in terms of planewaves
for coupling to planar surfaces, and so on~\cite{Rahi09:PRD}.

\begin{figure}[t]
\begin{centering}
\includegraphics[width=0.8\columnwidth]{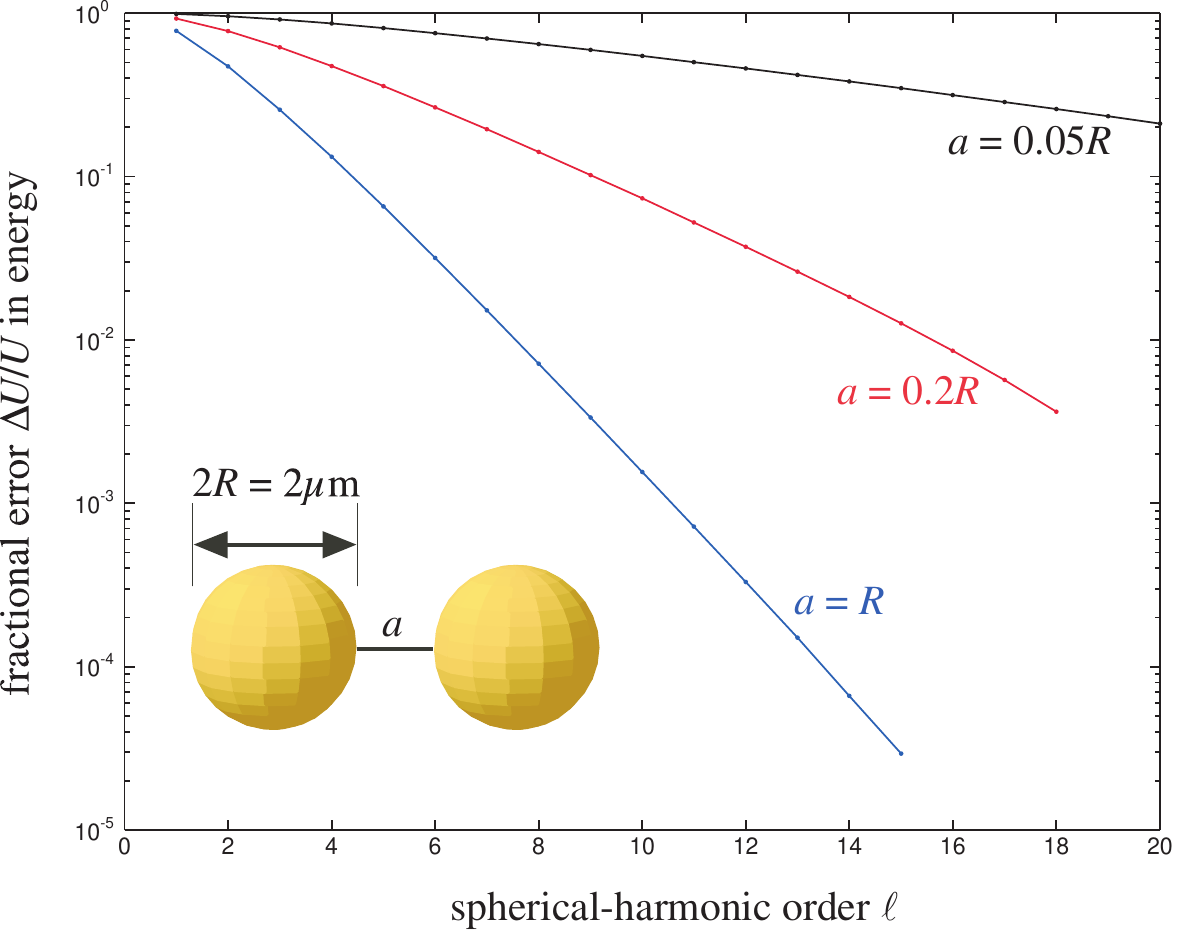}
\par\end{centering}

\caption{\label{fig:sphere-conv}Fractional error $\Delta U/U$ in the Casimir
interaction energy $U$ between two gold spheres of radius $R=1\,\mu$m,
for various surface-surface separations $a$, as a function of the
maximum spherical-harmonic order $\ell$ of the spectral path-integral
(scattering-matrix/multipole) method. The error converges exponentially
with $\ell$, but the exponential rate slows as $a/R$ shrinks. (Calculations
thanks to A.~Rodriguez.)}

\end{figure}

As was noted in Sec.~\ref{sub:Choices-of-basis}, such a spectral
basis can converge exponentially fast if there are no singularities
(e.g. corners) that were not accounted for analytically, and the method
can even lend itself to analytical study. Especially for cylinders
and spheres, the method is simple to implement and allows rapid exploration
of many configurations; the corresponding classical {}``multipole
methods'' are common in classical EM for cases where such shapes
are of particular interest~\cite{Yasumoto05}. On the other hand,
as the objects become less and less similar to the {}``natural''
shape for a given basis (e.g. less spherical for spherical harmonics),
especially objects with corners or cusps, the spectral basis converges
more slowly~\cite{KuoTi08}. Even for the interaction between two
spheres or a sphere and a plate, as the two surfaces approach one
another the multipole expansion will converge more slowly~\cite{Emig08-sphereplate,MaiaNeto08,EmigJa08}---conceptually,
a spherical-harmonic basis has uniform angular resolution all over
the sphere, whereas for two near-touching surfaces one would rather
have more resolution in the regions where the surfaces are close (e.g.
by using a nonuniform BEM mesh). This exponential convergence of a
spectral (spherical harmonic~\cite{Rahi09:PRD}) Casimir calculation
is depicted in Fig.~\ref{fig:sphere-conv} for the case of the Casimir
interaction energy $U$ between two gold spheres of radius $R=1\,\mu$m,
for various surface-to-surface separations $a$.%
 The error $\Delta U/U$ decreases exponentially with the maximum
spherical-harmonic order $\ell$ {[}corresponding to $N=4\ell(\ell+2)$
degrees of freedom for two spheres{]}, but the exponential rate slows
as $a/R$ decreases. (On the other hand, for small $a/R$ a perturbative
expansion or extrapolation may become applicable~\cite{Emig08-sphereplate}.)

\subsection{Boundary-element methods (BEMs)\label{sub:BEM-path}}

In a BEM, one meshes the interfaces, say into triangles, and uses
a set of localized basis functions $\vec{b}_{k}$ as discussed in
Sec.~\ref{sub:BEM-stress}. In this case, the interaction matrix
$A$ that arises in the path-integral formulation is\emph{ }exactly
the same as the interaction matrix that arises in classical BEM methods
(albeit at an imaginary frequency), and is the same as the matrix
$A$ that arises in a BEM stress-tensor approach as described in Sec.~\ref{sub:BEM-stress}.
The main difference, compared to the stress-tensor approach, lies
in how one \emph{uses} the matrix $A$: instead of solving a sequence
of linear equations to find the mean stress tensor $\langle\mathbf{T}\rangle$
at various points on a surface around an object, one computes $\log\det A$
or $\tr\left[A^{-1}\frac{\partial A}{\partial p_{i}}\right]$ to obtain
the energy~(\ref{eq:path-integral-energy-matrix}) or force~(\ref{eq:path-integral-force-matrix}).
We have demonstrated this approach for several three-dimensional geometries,
such as the crossed capsules of Fig.~\ref{fig:crossed-capsules}~\cite{ReidRo09}.

If one is using dense-matrix techniques, the advantage of this approach
over the stress-tensor technique seems clear~\cite{ReidRo09}: it
avoids the complication of picking a stress-integration surface and
an appropriate surface-integration technique, and allows the size
of the linear system to be easily reduced via blocking as in eq.~(\ref{eq:logdet-block}).
The situation is less clear as one moves to larger and larger problems,
in which dense-matrix solvers become impractical and one requires
an iterative method. In that case, computing $\tr\left[A^{-1}\frac{\partial A}{\partial p_{i}}\right]$
straightforwardly requires $N$ linear systems to be solved; if each
linear system can be solved in $O(N\log N)$ time with a fast solver
(as discussed in Secs.~\ref{sub:Solution-techniques} and~\ref{sub:BEM-stress}),
then the overall complexity is $O(N^{2}\log N)$ {[}with $O(N)$ storage{]},
whereas it is possible that the stress-tensor surface integral may
require fewer than $N$ solves. On the other hand, there may be more
efficient ways to compute the trace (or $\log\det$) via low-rank
approximations: for example, if the trace (or $\log\det$) is dominated
by a small number of extremal eigenvalues, then these eigenvalues
can be computed by an iterative method~\cite{bai00} with the equivalent
of $\ll N$ linear solves. The real-symmetric property of $A$, as
usual, means that the most favorable iterative methods can be employed,
such as a Lanczos or Rayleigh-quotient method~\cite{bai00}. Another
possibility might be sparse-direct solvers via a fast-multipole decomposition~\cite{GreengardGu09}.
The most efficient use of a fast $O(N\log N)$ BEM solver in Casimir
problems, whether by stress-tensor or path-integral methods, remains
an open question (and the answer may well be problem-dependent).

In the BEM approach with localized basis functions, the $\tr\left[A^{-1}\frac{\partial A}{\partial p_{i}}\right]$
expression for the force corresponds to a sum of a diagonal components
for each surface element, and in the exact limit of infinite resolution
(infinitesimal elements) this becomes an integral over the object
surfaces. Expressing the force as a surface integral of a quantity
related to Green's-function diagonals is obviously reminiscent of
the stress-tensor integration from Sec.~\ref{sub:Stress-tensor},
and it turns out that one can prove an exact equivalence using only
vector calculus~\cite{Reid09-PMCHW}. (At least one previous author
has already shown the algebraic equivalence of the stress tensor and
the derivative of the path-integral energy for forces between periodic
plates~\cite{Bimonte09}.)

\subsection{Hybrid BEM/spectral methods}

It is possible, and sometimes very useful, to employ a hybrid of the
BEM and spectral techniques in the previous two sections. One can
discretize a surface using boundary elements, and use this discretization
to solve for the scattering matrix $A_{kk}$ of each object in a spectral
basis such as spherical waves. That is, for any given incident spherical
wave, the outgoing field can be computed with BEM via (\ref{eq:BEM-abstract})
and then decomposed into outgoing spherical waves to obtain one row/column
of $A_{kk}$ at a time; alternatively, the multipole decomposition
of the outgoing wave can be computed directly from the multipole moments
of the excited surface currents $\vec{J}_{s}$~\cite{Jackson98}.
This approach appears to be especially attractive when one has complicated
objects, for which a localized BEM basis works well to express the
boundary conditions, but the interactions are only to be computed
at relatively large separations where the Casimir interaction is dominated
by a few low-order multipole moments. One can perform the BEM computation
once per object and re-use the resulting scattering matrix many times
via the analytical translation matrices, allowing one to efficiently
compute interactions for many rearrangements of the same objects and/or
for {}``dilute'' media consisting of many copies of the same objects~\cite{McCauleyZh10}.
(Essentially, this could be viewed as a form of low-rank approximation
of the BEM matrix, capturing the essential details relevant to moderate-range
Casimir interactions in a much smaller matrix.) Such a hybrid approach
is less attractive for closer separations, however, in which the increasing
number of relevant multipole moments will eventually lead to an impractically
large matrix to be computed.

\subsection{Eigenmode-expansion/RCWA methods}

\begin{figure}[t]

\begin{centering}
\includegraphics[width=0.7\columnwidth]{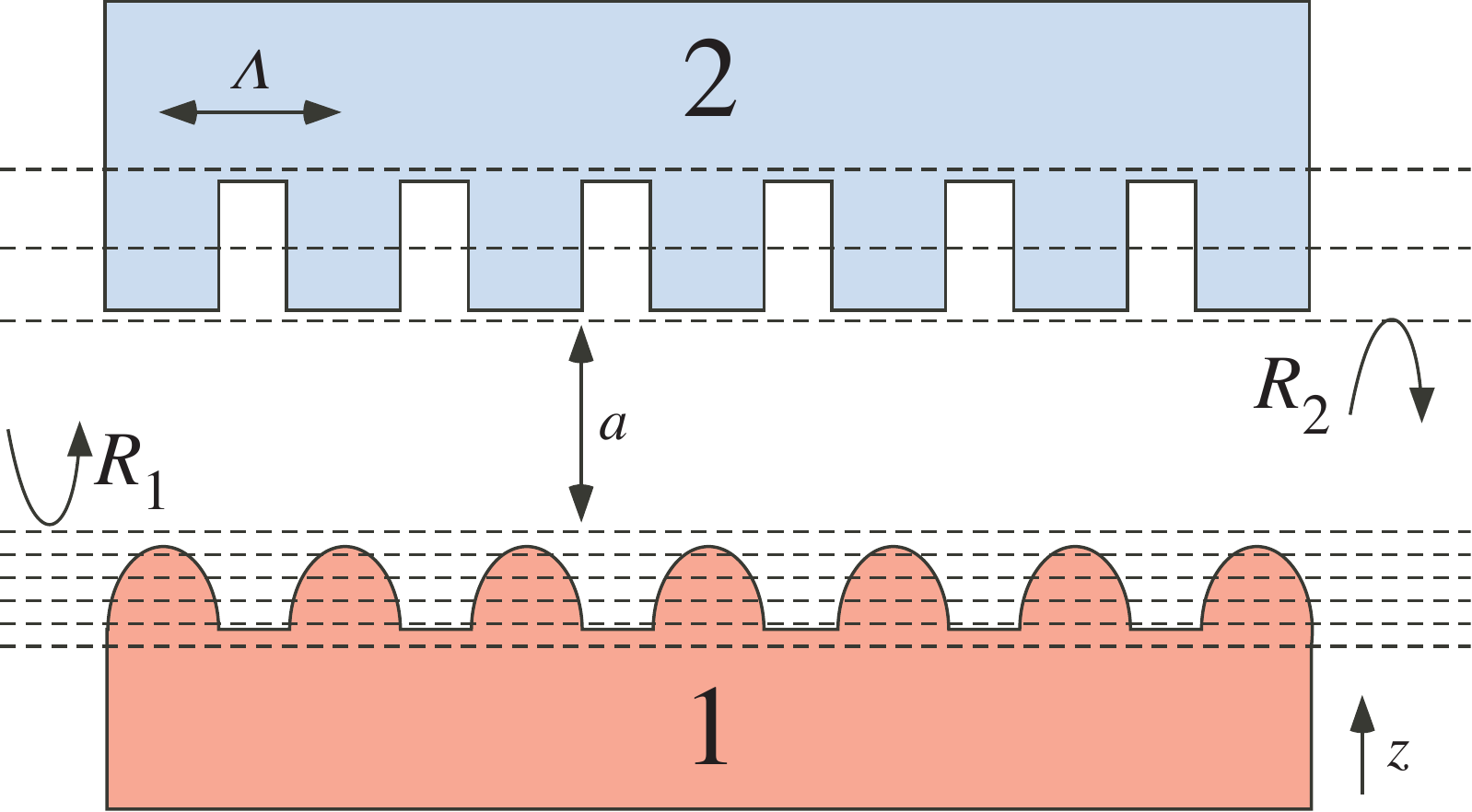}
\par\end{centering}

\caption{\label{fig:rcwa}Schematic problem for which eigenmode-expansion is
well suited: the interaction between two corrugated surfaces, with
period $\Lambda$. The Casimir problem reduces to computing the reflection
matrices $R_{1}$ and $R_{2}$ for each individual surface, in a planewave
basis. Eigenmode expansion works by expanding the field in each cross-section
(dashed lines) in the basis of eigenmodes of a $z$-invariant structure
with that cross-section, and then matching boundary conditions whenever
the cross-section changes.}

\end{figure}

Consider the case of the interaction between two corrugated surfaces
depicted in Fig~\ref{fig:rcwa}, separated in the $z$ direction.%
 From the scattering-matrix viewpoint, it is natural to consider scattering
off of each object by planewaves. In this case, the self-interaction
matrices $A_{11}^{-1}$ and $A_{22}^{-1}$ can be re-expressed in
terms of reflection matrices $R_{1}$ and $R_{2}$ for each surface,
relating the amplitudes of incident waves at some plane (dashed line)
above each surface to the reflected (specular and non-specular) planewave
amplitudes. The matrices $A_{12}$ and $A_{21}$ are replaced by a
diagonal matrix $D_{12}=D_{21}^{T}$ that relates the planewave amplitudes
at the planes for objects~1 and~2, separated by a distance $a$---at
real frequencies, this would be a phase factor, but at imaginary frequencies
it is an exponential decay as discussed below. This results in the
following expression for the Casimir interaction energy: \begin{equation}
U=\frac{\hbar c}{2\pi}\int_{0}^{\infty}\log\det\left[I-R_{2}D_{12}R_{1}D_{12}\right]d\xi.\label{eq:casimir-energy-R}\end{equation}
Alternatively, instead of viewing it as a special case of the T-matrix/scattering-matrix
idea~\cite{Rahi09:PRD}, the same expression can be derived starting
from an eigenmode-summation approach~\cite{Lambrecht09}.

The problem then reduces to computing the scattering of an incident
planewave off of a corrugated surface, with the scattered field decomposed
into outgoing planewaves. For this problem, one could use any of the
tools of computational EM (such as BEM, FD, and so on), but there
is a notable method that is often well-suited to the case of periodic
surfaces, especially periodic surfaces with piecewise-constant cross-sections\footnote{See also the chapter by A.~Lambrecht et~al.\ in this volume for additional discussion of Casimir interactions among periodic structures.}
(as in object~2 of Fig.~\ref{fig:rcwa}). This method is called
\emph{eigenmode expansion}~\cite{Willems95,Bienstman01,BienstmanBa02}
or \emph{rigorous coupled-wave analysis} (RCWA)~\cite{MoharamGr95,MoharamPo95},
or alternatively a \emph{cross-section} method~\cite{Katsenelenbaum98}.
RCWA has a long history because it is closely tied to semi-analytical
methods to study waveguides with slowly/weakly varying cross-sections~\cite{Marcuse74,Katsenelenbaum98}.
An analogous method was recently applied to Casimir problems~\cite{Lambrecht09}.
In RCWA, one computes reflection and scattering matrices at a given
frequency $\omega$ along some direction $z$ by expanding the fields
at each $z$ in the basis of the eigenmodes of the cross-section at
that $z$ (waves with $z$ dependence $e^{i\beta z}$ at a given $\omega$,
where $\beta$ is called the \emph{propagation constant} of the mode).
Along regions of uniform cross-section, the $z$ dependence $e^{i\beta z}$
of each mode is known analytically and no computation is required
(the mode amplitudes are multiplied by a diagonal propagation matrix
$D$). Regions of continuously varying cross-section are approximated
by breaking them up into a finite number of constant--cross-section
layers (as in object~1 of Fig.~\ref{fig:rcwa}). At any $z$ where
the cross-section changes, a change of basis is performed by matching
boundary conditions (the $xy$ components of the fields must be continuous),
yielding a \emph{transfer matrix} at that interface. All these transfer
and propagation matrices can then be combined to compute scattering/reflection
matrices for an entire structure.

The main difference here from classical RCWA computations is that
the modes are computed at imaginary frequencies $\xi$. As in Sec.~\ref{sub:imaginary-Green},
this actually simplifies the problem. At an imaginary frequency $\omega=i\xi$,
the modes of a given cross-section $\varepsilon(i\xi,x,y)$ and $\mu(i\xi,x,y)$
with $z$ dependence $e^{i\beta z}=e^{-\gamma z}$ ($\gamma=-i\beta$)
satisfy the eigenequation (for isotropic materials)~\cite{Johnson02:adiabatic,SkorobogatiyYa09}:\begin{multline}
\left(\begin{array}{cc}
\xi\varepsilon+\nabla_{xy}\times\frac{1}{\xi\mu}\nabla_{xy}\times{}\\
 & \xi\mu+\nabla_{xy}\times\frac{1}{\xi\varepsilon}\nabla_{xy}\times{}\end{array}\right)\left(\begin{array}{c}
\vec{E}_{xy}\\
\vec{H}_{xy}\end{array}\right)\\
=\gamma\left(\begin{array}{cccc}
 &  &  & 1\\
 &  & -1\\
 & -1\\
1\end{array}\right)\left(\begin{array}{c}
\vec{E}_{xy}\\
\vec{H}_{xy}\end{array}\right),\label{eq:waveguide-eig}\end{multline}
where the $xy$ subscript indicates a two-component vector with $xy$
(transverse) components. The operators on both the left- and right-hand
sides are real-symmetric, while the operator on the left-hand side
is positive-definite, and as a result the eigenvalues $\gamma$ are
purely real. This means that the propagation constants $\beta$ are
purely imaginary (all of the imaginary-frequency modes are evanescent
in $z$), and the analogues of incoming/outgoing waves are those that
are exponentially decaying towards/away from the surface. Moreover,
the numerical problem of solving for these eigenmodes in a given cross-section
reduces to a positive-definite generalized eigenvalue problem (a definite
matrix pencil~\cite{anderson99}), to which the most desirable numerical
solvers apply \cite{anderson99,bai00} (unlike the classical real-$\omega$
problem in which there are both propagating and evanescent modes because
the problem is indefinite~\cite{Johnson02:adiabatic,SkorobogatiyYa09}).
For homogeneous cross-sections (as in the space between the two objects),
the solutions are simply planewaves of the form $e^{ik_{x}x+ik_{y}y-\gamma z+\xi t}$,
where for vacuum $\gamma=\pm\sqrt{|\vec{k}_{xy}|^{2}+\xi^{2}/c^{2}}$.

For sufficiently simple cross-sections, especially in two-dimensional
or axisymmetric geometries, it is possible to solve for the modes
analytically and hence obtain the scattering matrices, and this is
how the technique was first applied to the Casimir problem~\cite{Lambrecht09}.
For more general geometries, one can solve for the modes numerically
by a variety of techniques, such as by a transfer-matrix method in
two dimensions~\cite{Bienstman01} or by a planewave expansion (in
the $xy$ cross-section) in three dimensions~\cite{MoharamGr95}.
Of course, one truncates to a finite number of modes via some cutoff
$|\gamma|$ (which follows automatically from discretizing the cross-section
in a finite grid, for example), and convergence is obtained in the
limit as this cutoff increases. Given a basis of eigenmodes with some
cutoff, the process of constructing the scattering/reflection matrices
is thoroughly discussed elsewhere~\cite{Willems95,Bienstman01,BienstmanBa02,MoharamGr95,MoharamPo95},
so we do not review it here.

The strength of RCWA is that regions of uniform cross-section are
handled with at most a 2d discretization of the cross-section, independent
of the thickness of the region, so very thick or very thin layers
can be solved efficiently. The main limitation of RCWA methods is
that the transfer matrices (and the resulting reflection matrices
$R_{1}$ and $R_{2}$) are dense $N\times N$ matrices, where $N$
is the number of modes required for convergence. If $N$ is large,
as in complicated three-dimensional structures, the problem can quickly
become impractical because of the $O(N^{2})$ storage and $O(N^{3})$
computation requirements. The most favorable case is that of periodic
structures with relatively simple unit cells, in which case the problem
can be reduced to that of computing the modes of each periodic unit
cell (with Bloch-periodic boundary conditions) as discussed below,
and RCWA can then be quite practical even in three dimensions. Non-periodic
structures, such as compact objects, can be handled by perfectly matched
layer (PML) absorbing boundaries~\cite{BienstmanBa02}, albeit at
greater computational cost because of the increased cross-section
size.

\section{Periodicity and other symmetries}

In this section, we briefly discuss the issue of periodicity and other
symmetries, which can be exploited to greatly reduce the computational
effort in Casimir calculations just as for classical EM calculations.

If a structure is periodic in the $x$ direction with period $\Lambda$,
as in Fig.~\ref{fig:rcwa}, the problem simplifies considerably because
one can reduce the computation to a single unit cell of thickness
$\Lambda$. In particular, one imposes Bloch-periodic boundary conditions---the
fields at $x=\Lambda$ equal the fields at $x=0$ multiplied by a
phase factor $e^{ik_{x}\Lambda}$---and computes the Casimir energy
or force for each Bloch wavevector $k_{x}$ separately, then integrates
the result over $k_{x}$ via $\int_{-\pi/\Lambda}^{\pi/\Lambda}(\cdots)dk_{x}$.
This can be derived in a variety of ways, for example by applying
Bloch's theorem~\cite{JoannopoulosJo08-book} to decompose the eigenmodes
into Bloch-wave solutions for each $k_{x}$, or by expanding the delta
functions of the fluctuation--dissipation approach in a Fourier series~\cite{Rodriguez07:PRA}.
More generally, for any periodic unit cell, one can perform the Casimir
energy/force computation for the unit cell with Bloch periodic boundaries
and then integrate the Bloch wavevector $\vec{k}$ over the irreducible
Brillouin zone (multiplied by the volume ratio of the Brillouin zone
and the irreducible Brillouin zone).

The specific case of continuous translational symmetry, say in the
$x$ direction, corresponds $\Lambda\to0$ and one must integrate
over all $k_{x}$ (the Brillouin zone is infinite). Certain additional
simplifications apply in the case of a perfect-metal structure with
continuous translational symmetry, in which case the fields decompose
into two polarizations and the $k$ integration can be performed implicitly~\cite{Rodriguez07:PRA}.

Rotational symmetry can be handled similarly: the fields can be decomposed
into fields with $e^{im\phi}$ angular dependence, and the total force
or energy is the sum over all integers $m$ of the contributions for
each $m$~\cite{McCauleyRo10:PRA}. More generally, the Casimir contributions
can be decomposed into a sum of contributions from irreducible representations
of the symmetry group of the structure (e.g. all eigenmodes can be
classified into these representations~\cite{Inui96,Tinkham03});
translational and rotational symmetries are merely special cases.
As another example, in a structure with a mirror symmetry one could
sum even- and odd-symmetry contributions (in fact, this is the underlying
reason for the TE/TM polarization decomposition in two dimensions~\cite{JoannopoulosJo08-book}).

\section{Nonzero-temperature corrections\label{sec:finite-temperature}}

In the preceding sections, we discussed only the computation of Casimir
interactions at zero temperature $T=0^+$. However, the modification
of any imaginary-frequency expression for a Casimir interaction from
$T=0$ to $T>0$ is almost trivial: one simply performs a sum instead
of an integral. If the $T=0$ interaction (energy, force, etc.) is
expressed as an integral $\int_{0}^{\infty}C(\xi)\, d\xi$ of some
contributions $C(\xi)$ at each imaginary frequency $\xi$, then the
$T>0$ interaction is well known to be simply~\cite{Lifshitz80}:
\begin{equation}
\int_{0}^{\infty}C(\xi)d\xi\to\frac{2\pi k_{\mathrm{B}}T}{\hbar}\sideset{}{'}\sum_{n=0}^{\infty}C\left(\frac{2\pi k_{\mathrm{B}}T}{\hbar}n\right),\label{eq:nonzero-T}\end{equation}
where $k_{\mathrm{B}}$ is Boltzmann's constant and $\sum'$ indicates
a sum with weight $\tfrac{1}{2}$ for the $n=0$ term. The frequencies
$\xi_{n}=2\pi k_{\mathrm{B}}Tn/\hbar$ are known as Matsubara frequencies,
and the corresponding (imaginary) Matsubara wavelengths are $\lambda_{n}=2\pi/\xi_{n}=\lambda_{T}/n$
where $\lambda_{T}=\hbar/k_{\mathrm{B}}T$. The conversion of the
$T=0$ integral into a summation can be derived in a variety of ways,
most directly by considering thermodynamics in the Matsubara formalism~\cite{Lifshitz80}.
Physically, this arises from the $\coth(\hbar\omega/2kT)$ Bose--Einstein
distribution factor that appears in the fluctuation--dissipation expressions~(\ref{eq:Ecorr-real})
for nonzero temperatures. When the contour integration is performed
over $\omega$, the $\coth$ introduces poles at $\hbar\omega/2kT=i\pi n$
that convert the integral into a sum via the residue theorem (with
the $n=0$ residue having half weight because it lies on the real-$\omega$
axis)~\cite{Lamoreaux05}. As explained in Sec.~\ref{sub:imaginary-Green},
some care must be applied in evaluating the $n=0$ term because of
the well known singularity of Maxwell's equations at $\omega=0$ (where
the $\vec{E}$ and $\vec{H}$ fields decouple), and one may need to
take the limit $\xi\to0^{+}$ (although there is some controversy
in the unphysical case of perfect metals~\cite{Hoye05})

Mathematically, the sum of eq.~(\ref{eq:nonzero-T}) is exactly the
same as a trapezoidal-rule approximation for the $T=0$ integral,
with equally spaced abscissas $\Delta\xi=2\pi/\lambda_{T}$~\cite{boyd01:book,Press92,RodriguezWo10:arxiv}.
Thanks to the $O(\Delta\xi^{2})$ convergence of the trapezoidal rule~\cite{boyd01:book},
this means that the $T>0$ result is quite close to the $T=0$ result
unless $C(\xi)$ varies rapidly on the scale of $2\pi/\lambda_{T}$.
In particular, suppose that $C(\xi)$ varies on a scale $2\pi/a$
, corresponding to some lengthscale $a$ in the problem (typically
from a surface--surface separation). In that case, assuming $C(\xi)$
has nonzero slope%
\footnote{If $C(\xi)$ has zero slope at $\xi=0^{+}$, then the trapezoidal
rule differs from the integral by $O(\Delta\xi^{4})$ or less, depending
upon which derivative is nonzero at $\xi=0^{+}$~\cite{boyd01:book}.%
} at $\xi=0^{+}$ (typical for interactions between realistic metal surfaces),
then the nonzero-$T$ correction should be of order $O(a^{2}/\lambda_{T}^{2})$.
At room temperature ($T=300$~K), $\lambda_{T}\approx7.6\,\mu$m,
and the temperature corrections to Casimir interactions are typically
negligible for submicron separations~\cite{Lifshitz80,milton04}.
On the other hand, it is possible that careful material and geometry
choices may lead to larger temperature effects~\cite{RodriguezWo10:arxiv}.
There is also the possibility of interesting effects in nonequilibrium
situations (objects at different temperatures)~\cite{Najafi04,Obrecht07},
but such situations are beyond the scope of this review.

\section{Concluding remarks}

The area of numerical Casimir computations remains rich with opportunities.
Relatively few geometry and material combinations have as yet been
explored, and thus many newly answerable questions remain regarding
the ways in which Casimir phenomena can be modified by exploiting
the degrees of freedom available in modern nanofabrication. In the
regime of computational techniques, while several effective methods
have already been proposed and demonstrated, the parallels with computational
electromagnetism lead us to anticipate ongoing improvements and developments
for some time to come. The same parallels also caution against any
absolute {}``rankings'' of the different approaches, as different
numerical techniques have always exhibited unique strengths and weaknesses
in both theory and practice. And because computer time is typically
much less expensive than programmer time, there is something to be
said for methods that may be theoretically suboptimal but are easy
to implement (or are available off-the-shelf) for very general geometries
and materials. Nor is the value of analytical and semi-analytical
techniques diminished, but rather these approaches are freed from
the tedium of hand computation to focus on more fundamental questions.

\section*{Acknowledgements}

This work was supported in part by the Army Research Office through
the ISN under contract W911NF-07-D-0004, by the MIT Ferry Fund, and
by the Defense Advanced Research Projects Agency (DARPA) under contract
N66001-09-1-2070-DOD. We are especially grateful to our students,
A.~W.~Rodriguez, A.~P.~McCauley, and H.~Reid for their creativity
and energy in pursuing Casimir simulations. We are also grateful to
our colleagues F. Capasso, D.~Dalvit, T.~Emig, R.~Jaffe, J.~D.~Joannopoulos,
M.~Kardar, M.~Levin, M.~Lon\v{c}ar, J.~Munday, S.~J.~Rahi, and
J.~White, for their many suggestions over the years.

\bibliographystyle{spmpsci}

\end{document}